\documentclass[useAMS,referee,usenatbib]{biom}

\def\bSig\mathbf{\Sigma}

\usepackage{graphicx} 
\usepackage{amsfonts,amssymb, amsmath} 
\usepackage{mathtools}
\usepackage{upgreek}
\usepackage{color}
\usepackage{subfigure}
\usepackage{xcolor} 
\usepackage{subfigure}
\usepackage{algorithm}
\usepackage{algpseudocode}
\usepackage{booktabs}
\usepackage[colorlinks=true, linkcolor=blue, citecolor=blue, urlcolor=blue]{hyperref}

\title{Asymptotic Distribution of Robust Effect Size Index}

\author{Xinyu Zhang$^{1,*}$\email{xinyu.zhang@vanderbilt.edu, simon.vandekar@vumc.org}, 
Rachael Muscatello$^{2}$,  
Megan Jones$^{1}$, Blythe Corbett$^{2}$, and Simon Vandekar$^{1,3,*}$ \\
$^{1}$Department of Biostatistics, Vanderbilt University \\
$^{2}$Department of Psychiatry \& Behavioral Sciences, Vanderbilt University Medical Center \\
$^{3}$Vanderbilt University Medical Center}

\begin{document}


\date{{\it Received xxx} 2026. {\it Revised xxx} 2026.  {\it
Accepted xxx} 2026.}



\pagerange{\pageref{firstpage}--\pageref{lastpage}} 
\volume{xx}
\pubyear{xxxx}
\artmonth{xxx}


\doi{10.1111/j.1541-0420.2005.00454.x}


\label{firstpage}


\begin{abstract}

The Robust Effect Size Index (RESI) is a recently proposed standardized effect size to quantify association strength across models. However, its confidence interval construction has relied on computationally intensive bootstrap procedures. We establish a general theorem for the asymptotic distribution of the RESI using a Taylor expansion that accommodates a broad class of models. Simulations under various linear and logistic regression settings show that RESI and its CI have smaller bias and more reliable coverage than commonly used effect sizes such as Cohen’s $d$ and $f$. Combining with robust covariance estimation yields valid inference under model misspecification. We use the methods to investigate associations of depression and behavioral problems with sex and diagnosis in Autism spectrum disorders and demonstrate that the asymptotic approach achieves up to a 50-fold speedup over the bootstrap. Our work provides a scalable and reliable alternative to bootstrap inference, greatly enhancing the applicability of RESI to high-dimensional studies. 

\end{abstract}

%

\begin{keywords}
Semiparametric; Generalized linear models; Autism spectrum disorders; Hypothesis testing.
\end{keywords}


\maketitle


%

\section{Introduction}\label{sec:intro}

Effect size indices are parameters that quantify the magnitude of a phenomenon in a scale that is independent of sample size \citep{kelley_effect_2012}. While $p$-values quantify evidence against a null hypothesis, they do not indicate the practical significance of an effect and are highly sensitive to sample size \citep{wasserstein_asas_2016, wasserstein_moving_2019, trafimow_null_2017, kafadar_editorial_2021}.  Effect size estimates are a critical tool in study design, power analysis, and meta-analysis \citep{cohen_statistical_1988, chinn_simple_2000, morris_combining_2002}. 
In bio-behavioral sciences, they are particularly valuable due to the broad array of measurement instruments used across studies that have different scoring distributions, making unstandardized metrics (e.g., raw mean differences between diagnostic groups) difficult to compare \citep{cohen_statistical_1988, hedges_statistical_1985, wasserstein_moving_2019}. 
In such settings, standardized effect sizes are essential for replicability: they remain interpretable across heterogeneous measures and support quantitative replication and integration of results \citep{kang_study_2024, cumming_new_2014, lakens_equivalence_2017}.

As a representative example, characterizing associations of depression and behavioral problems through development in individuals with autism spectrum disorders (ASDs) is important to prioritize, develop, and time specialized interventions.
Depression can be measured with the Children's Depression Inventory (CDI) \citep{kovacs_cdi_1985}, while behavioral problems are typically summarized using a T-score from the Child Behavioral Checklist (CBCL) \citep{achenbach_manual_2001} and often analyzed continuously or as a binary outcome (by dichotomizing the T-score).
Effect sizes estimates are valuable in this context to compare association strengths across studies \citep{lord_autism_2020, lombardo_big_2019, happe_time_2006,corbett_developmental_2021}. 
Due to the heterogeneity of ASDs, the distributions of these psychiatric measures are heteroskedastic and skewed \citep{gotham_standardizing_2009}, which complicates effect size estimation and inference. 
As a result, ASD studies provide an ideal testbed for evaluating effect size methods that aim to be robust, comparable across diverse outcomes, and applicable beyond Gaussian models.

Traditional standardized effect sizes, such as Cohen’s $d$ (for pairwise comparisons) and Cohen’s $f$ (for analysis of variance (ANOVA)) \citep{cohen_statistical_1988}, are limited in several ways. First, their estimates and confidence intervals (CIs) rely on strict assumptions of normality and homoskedasticity.
Second, these indices are defined within a linear-model framework and lack unified extensions to generalized linear model (GLM) \citep{cohen_statistical_1988, olejnik_generalized_2003}. 
Although GLMs provide model-specific measures such as odds ratios or rate ratios, these are not standardized effect sizes and do not offer a common scale \citep{vandekar_robust_2020}.
The Robust Effect Size Index (RESI) establishes a generalized framework based on estimating equations, offering broad applicability across statistical models, robustness to model misspecification via heteroskedasticity consistent variance estimation, and the ability to accommodate nuisance covariates \citep{vandekar_robust_2020}. 
However, inference for the RESI remains a theoretical challenge because the estimator's distribution depends on the joint distribution of parameter estimators and their covariance estimator.
Current RESI inference relies on bootstrap procedures \citep{kang_accurate_2023}, is implemented in the \textit{RESI} R package \citep{jones_resi_2025}. 
Bootstrapping imposes substantial computational costs, particularly for large-scale applications ($n \ge 500$) or high-dimensional studies \citep{kang_accurate_2023, zhang_semiparametric_2025}.
These considerations motivate the development of a direct asymptotic approximation that offers a more scalable and theoretically grounded alternative for RESI inference.

In this paper, we derive the asymptotic distribution of the RESI estimator from the estimating equation structure of the underlying model. By applying the multivariate Delta method to the joint distribution of the model parameters and their robust sandwich covariance implied by the estimating equations, we establish a general theorem for the RESI estimator's asymptotic behavior. This approach accommodates a wide class of models without assuming data normality.
To address the distributional irregularity of unsigned effect sizes near zero, we propose a truncated interval construction method to ensure valid interval estimation at the boundary.
This analytic formulation leads to a new CI estimation algorithm that substantially avoids computational burden and boundary issues associated with bootstrap methods.
We apply this approach to two ASD-related settings. One examines diagnostic differences in developmental trajectories of depressive symptoms measured by CDI, and the other uses large-scale data from the SPARK study to assess sex differences in developmental patterns of behavioral severity measured by CBCL.

\section{Method}

\subsection{Estimating Equation Notation}
\label{sec:estimatingeq}

Let $W = (W_1, \cdots, W_n)$ be independent random vectors representing the data for $n$ subjects. We define the estimating equation
\begin{equation}
\label{eq:estimatingEquation}
    \Psi_n\left(\theta^*; W\right) = \frac{1}{n} \sum_{i=1}^n \psi \left(\theta^*; W_i \right),
\end{equation}
where $\theta^* \in \Theta$ is a generic parameter value and $\psi(\theta^*, \cdot) \in \mathbb{R}^m$ is a known estimating function. $\psi$ covers the standard score-type estimating equations, including those arising from likelihood-based models and from quasi-likelihood formulations such as GEE \citep{liang_longitudinal_1986}.  

We denote the true parameter vector by $\theta = (\alpha, \beta) \subset \mathbb{R}^m$, where $\alpha \in \mathbb{R}^{m_0}$ is a vector of nuisance parameters including intercept and possibly a dispersion parameter $\phi$ (if applicable), $\beta \in \mathbb{R}^{m_1}$ is the vector of target parameters, and $m = m_0 + m_1$. We define $\theta$ and its consistent estimator $\hat{\theta}$ as the solutions to the population and sample estimating equation $\mathbb{E} \left\{\psi(\theta^*; W) \right\} = 0$ and $\Psi_n(\theta^*; W) = 0$, respectively.
Under regularity conditions in the Supplement Section \ref{sec:regularity} \citep{van_der_vaart_asymptotic_2000, boos_essential_2013}, the $\hat{\theta}$ is consistent and asymptotically normal
\citep{white_heteroskedasticity-consistent_1980}
\begin{equation*} 
\sqrt{n}(\hat{\theta} - \theta) \xrightarrow{d} \mathcal{N}(0, \Sigma_{\theta}), 
\end{equation*}
where the asymptotic ``robust" covariance matrix takes the sandwich form 
\begin{equation}
\label{eq:robustCov}
\Sigma_\theta = A_\theta^{-1}B_\theta A_\theta^{-1},
\end{equation} 
where $A_{\theta}$ and $B_{\theta}$ are $m \times m$ matrices \citep{van_der_vaart_asymptotic_2000}
\begin{equation*}
A_{\theta} = -\mathbb{E} \left\{ \frac{\partial \psi(\theta^*; W)}{\partial \theta^{*\top}}\Big|_{\theta^* = \theta} \right\} = -\mathbb{E}\left\{\psi'(\theta; W)\right\}, \quad
B_{\theta} = \mathbb{E} \left\{ \psi(\theta; W) \psi(\theta; W)^\top \right\}.
\end{equation*}
This heteroskedasticity-consistent covariance ensures valid inference even if $\psi$ is not based on the log-likelihood \citep{white_heteroskedasticity-consistent_1980, huber_robust_1964}. 
When the estimating equation is equal to the derivative of the correctly specified log-likelihood, then $A_{\theta} = B_\theta$ and the covariance is
\begin{equation}\label{eq:paramCov}
    \Sigma_\theta = A_\theta^{-1}.
\end{equation}

The asymptotic covariance matrix $\widehat \Sigma_\theta$ can be estimated with plug-in estimators. Since $\widehat A_\theta$ and $\widehat B_\theta$ are consistent estimators of $A_\theta$ and $B_\theta$, the sandwich estimator $\widehat \Sigma_\theta = \widehat A_\theta^{-1}\widehat B_\theta \widehat A_\theta^{-1}$ is consistent by the continuous mapping theorem \citep{van_der_vaart_asymptotic_2000}. Specifically,
\begin{equation}
    \widehat{A}_{\theta} = -\frac{1}{n}\sum_{i=1}^n \psi'(\hat\theta; W_i), \quad 
    \widehat{B}_{\theta} = \frac{1}{n} \sum_{i=1}^n \psi(\hat{\theta}; W_i) \psi(\hat{\theta}; W_i)^\top.
    \label{eq:covEst}
\end{equation}

\subsection{The Robust Effect Size Index}\label{sec:resi}

The RESI is based on the Wald-type test statistic for testing the null hypothesis $ H_0: \beta = \beta_0$, where $\beta_0 \in \mathbb{R}^{m_1}$ is a fixed reference vector (typically $\mathbf{0}$). The statistic is given by
\begin{equation*}\label{eq:t2stat}
T^2 = n\left(\hat\beta - \beta_0\right)^\top \widehat\Sigma_\beta^{-1} \left(\hat\beta - \beta_0\right), 
\end{equation*}
where $\widehat\Sigma_\beta$ denotes the $m_1 \times m_1$ submatrix of the sandwich covariance estimator  $\widehat\Sigma_\theta$.
Assuming known variance (along with the regularity conditions in Supplement Section \ref{sec:regularity}), $T^2$ asymptotically follows a chi-squared distribution with $m_1$ degrees of freedom, with non-centrality parameter $
\lambda = n(\beta-\beta_0)^\top\Sigma_{\beta}^{-1}(\beta-\beta_0)$ (\citep[Theorems 5.21 and 5.23]{van_der_vaart_asymptotic_2000}; \citep[Appendix]{vandekar_robust_2020}). With the estimated variance, the distribution deviates from the theoretical non-central chi-square approximation because estimation introduces additional variability \citep{kang_accurate_2023}. 

The RESI, denoted by $S_\beta$, is defined as the square root of the non-centrality parameter $\lambda$ normalized by the sample size $n$:
\begin{equation}\label{eq:resi}
    S_\beta = \sqrt{(\beta-\beta_0)^\top\Sigma_{\beta}^{-1}(\beta-\beta_0)}.
\end{equation}
This definition yields a unitless effect size metric that is invariant to sample size.
When $m_1=1$, \cite{kang_accurate_2023} and \cite{jones_resi_2025} introduced a signed definition of the parameter,
$$S_\beta = (\beta-\beta_0)/\sigma_{\beta},$$
where $\sigma_\beta$ is the standard deviation of $\sqrt n \hat\beta$.
The RESI parameter depends subtly on features of the model and study design through $\Sigma_{\beta}$.

We consider the following estimators of $S_\beta$ proposed by \cite{jones_resi_2025}.
\begin{enumerate}
\item An unsigned estimator based on the T$^2$ statistic:
\begin{equation}\label{eq:shat_chisq}   
\widehat S_\beta = \left\{\max\left(0, \frac{T^2-m_1}{n}\right)\right\}^{1/2}.
\end{equation}


\item A signed estimator based on the Z statistics (for $m_1=1$):
\begin{equation}\label{eq:scheck_z}
    \check S_\beta = \frac{Z}{\sqrt{n}},
\end{equation}
where $Z = \frac{\hat\beta - \beta_0}{\widehat{se}(\hat \beta)}$ is the standard Z statistic. 
 

\item A scaled form of the T$^2$ statistics:
\begin{equation}\label{eq:stilde}
 \widetilde S_\beta = \left(\frac{T^2}{n}\right)^{1/2}.
\end{equation}
\end{enumerate}

These estimators have robust and parametric versions, which differ in whether the sandwich or model-based covariance matrix in \eqref{eq:robustCov} or \eqref{eq:paramCov} is used to estimate $\Sigma_\beta$ via the plug-in estimators \eqref{eq:covEst}.

There are two additional estimators based on the F and t statistics and used in linear models, which are asymptotically equivalent to $\widehat{S}_\beta$ and $\check{S}_\beta$ \citep{jones_resi_2025}. Further details about their asymptotic equivalence can be found in Supplement Section \ref{sec:asy_equiv_dist}.

We use estimators \eqref{eq:shat_chisq} and \eqref{eq:scheck_z} as point estimates and derive the asymptotic sampling distribution of \eqref{eq:stilde} to construct confidence intervals. Crucially, under the alternative hypothesis, all estimators are asymptotically equivalent up to order $O_p(n^{-1/2})$ (see Supplement Section \ref{sec:asy_equiv_dist}).

\subsection{The RESI Estimator Asymptotic Distribution}
\label{sec:theorem}

We use the multivariate delta method to derive the asymptotic variance of the RESI estimator. We view the RESI parameter $S_\beta$ as a function of $\theta$ directly and indirectly through $A_\theta$ and $B_\theta$.

The asymptotic covariance matrix of the estimators $\sqrt{n}(\hat \beta - \beta)$ is given by
\begin{equation*}
\begin{aligned}
    \Sigma_\beta &=  L^\top \left(A^{-1}_\theta B_\theta A^{-1}_\theta\right) L,
\end{aligned}
\end{equation*}
where $L \in \mathbb{R}^{m_1 \times m}$ is a linear constrast matrix such that $\beta =L \theta$.
The RESI for $\beta$ based on \eqref{eq:resi} is
\begin{equation*}
S_\beta(\theta) = \left\{\theta^\top L^\top \left(L A^{-1}_\theta B_\theta A^{-1}_\theta L^\top\right)^{-1}  L \theta\right\}^{1/2}.
\end{equation*}
We study the asymptotic behavior of the plug-in estimator $\widetilde{S}_\beta = S_\beta(\hat{\theta})$, where $A_\theta$ and $B_\theta$ are replaced by the estimators in \eqref{eq:covEst}.

\begin{theorem}[Asymptotic Distribution of $\widetilde S_\beta$]\label{thm:theorem1}
Let $\theta = (\alpha, \beta) \subset \mathbb{R}^m$ be the true parameter, and let $\hat{\theta}$ be the solution to Equation \eqref{eq:estimatingEquation}. Assume that
\begin{enumerate}
  \item[(A1)] $\hat\theta \xrightarrow{p} \theta$ and $\sqrt{n}(\hat{\theta} - \theta) \xrightarrow{d} \mathcal{N}(0, \Sigma_{\theta})$ based on the regularity conditions in Supplement Section \ref{sec:regularity};
  \item[(A2)] $S_\beta(\theta)$ is continuously differentiable in a neighborhood of the true parameter $\theta$.
\end{enumerate}
Then
\begin{equation}\label{eq:asympDist}
\sqrt{n}(\widetilde{S}_\beta - S_\beta) \xrightarrow{d} \mathcal{N}(0, \sigma_S^2),
\end{equation}
where the asymptotic variance is given by:
\begin{equation}
\label{eq:resiVar}
\sigma_S^2 = \left( \frac{d S_\beta}{d \theta} \right)^\top \Sigma_\theta \left( \frac{d S_\beta}{d \theta} \right).
\end{equation}
Here, $\frac{d S_\beta}{d \theta}$ denotes the total derivative of the function $S_\beta(\theta)$ with respect to $\theta$,
accounting for the dependence of matrices $A_\theta$ and $B_\theta$ on $\theta$.
$$\frac{d S_\beta}{d \theta}
= \frac{\partial S_\beta}{\partial \theta}
+ \left(\frac{d \mathrm{vec}(A_\theta)}{d \theta}\right)^\top \frac{\partial S_\beta}{\partial \mathrm{vec}(A_\theta)}
+ \left(\frac{d \mathrm{vec}(B_\theta)}{d \theta}\right)^\top \frac{\partial S_\beta}{\partial \mathrm{vec}(B_\theta)},$$
where $d$ denotes the total derivative, $\partial$ is the partial derivative, and $\mathrm{vec}(\cdot)$ flattens matrices into column-major and $\Sigma_\theta$ is the robust covariance as defined in \eqref{eq:robustCov}.
See Supplement Section \ref{sec:derivatives} and \ref{sec:proof_thm1} for formulas of the derivatives and the proof of the theorem.
\end{theorem}

When the parametric covariance of $\sqrt{n}(\hat \theta -\theta)$ is used to estimate $\widetilde S_\beta$, then the derivative is simplified.

\begin{corollary}
\label{cor:paramEst}
Assume conditions as in Theorem \ref{thm:theorem1} and that $\widetilde S_\beta = \left\{\hat \theta^\top L^\top \left(L \widehat A^{-1}_\theta L^\top\right)^{-1}  L \hat \theta\right\}^{1/2}$, the results in Equations \eqref{eq:asympDist} and \eqref{eq:resiVar} hold, where
$$
\frac{dS_\beta}{d\theta}
=
\frac{\partial S_\beta}{\partial \theta}
+
\left(\frac{d\,\mathrm{vec}(A_\theta)}{d\theta}\right)^\top
\frac{\partial S_\beta}{\partial \mathrm{vec}(A_\theta)},
$$
and $\Sigma_\theta$ is the robust covariance as defined in \eqref{eq:robustCov}.
\end{corollary}

Finally, if the parametric variance is used and $\psi$ is proportional to the derivative of the correctly specified log-likelihood, then the RESI asymptotic distribution is further simplified.

\begin{corollary}
Assume conditions as in Corollary \ref{cor:paramEst} and that $\psi$ is proportional to the derivative of the correctly specified log-likelihood, then
$$\sqrt{n}(\tilde{S}_\beta - S_\beta) \xrightarrow{d} \mathcal{N}(0, \sigma_S^2),$$
where the asymptotic variance, $\sigma^2_S$ is as in Theorem \ref{thm:theorem1}, with the model-based covariance $\Sigma_\theta$ as defined in Equation \eqref{eq:paramCov}.
\end{corollary}

The asymptotic variance $\sigma^2_S$ is well-defined for $S_\beta > 0$ in the Theorem \ref{thm:theorem1} and we show that $\sigma_S^2 \to 1$ as $S_\beta \to 0$ (Supplement Section \ref{sec:limit_variance}).



\subsection{Confidence Interval Construction}

For the unsigned estimator $\widehat S_{\beta}$, the parameter space is restricted to $[0, \infty)$. 
The asymptotic behavior differs on the boundary $S_{\beta}=0$.
The limiting distribution under the null is a censored chi-square distribution $\sqrt{\max(0, \chi^2_{m_1} - m_1)}$ (see Supplement Section \ref{sec:asy_equiv_dist}).
Consequently, standard intervals constructed near the boundary often yield logically inconsistent lower bounds (i.e., $S_L < 0$) or poor coverage probabilities.
To address this, we propose a truncated CI procedure that adjusts the bounds to respect the non-negative parameter space while accounting for the probability mass at the boundary (Algorithm \ref{alg:truncated_ci}).

\begin{algorithm}[h]
\caption{Truncated Confidence Interval Construction for Unsigned RESI}
\label{alg:truncated_ci}
\begin{algorithmic}[1]
\Require RESI estimate $\widehat{S}_\beta$,  sample size $n$, standard error $SE = \widehat{\sigma}_S/\sqrt{n}$, degrees of freedom $m_1$,  significance level $\alpha$
\Ensure Confidence Interval $[S_L, S_U]$.

\State Compute the standard Wald $1-\alpha$ confidence interval: 
$$[S_L, S_U] := \left[\widehat{S}_\beta - z_{1-\alpha/2} \cdot SE,\quad \widehat{S}_\beta + z_{1-\alpha/2} \cdot SE\right]$$

\If{$S_L \le 0$}
    \State $S_L := 0$
    \State Compute the right-tail probability of $\widehat S_\beta$ under the null: $\gamma = \mathbb{P}_{\text{H}_0}(\widehat S_\beta > s)$.
    
    \If{$\gamma < \alpha/2$}
        \State Adjust the upper bound quantile to maintain coverage:
        $$[S_L, S_U] := \left[0,\quad \widehat S_\beta + z_{1-(\alpha - \gamma)} \cdot SE\right]$$
    \Else
        \State Construct the standard one-sided $1-\alpha$
        interval:
        $$[S_L, S_U] := \left[0,\quad \widehat S_\beta + z_{1-\alpha} \cdot SE\right]$$
    \EndIf
\EndIf

\State \Return $[S_L, S_U]$
\end{algorithmic}
\end{algorithm}

For the signed estimator $\check S_{\beta}$, the parameter space is the entire real line ($\mathbb{R}$). The asymptotic distribution is as stated in Theorem \ref{thm:theorem1},
$\sqrt{n}(\check{S}_{\beta} - S_{\beta}^\pm) \xrightarrow{d} \mathcal{N}(0, \sigma_S^2)$, where $S_\beta^\pm = sgn(\beta-\beta_0)S_\beta$ and $sgn$ is the sign function. Under the null $S_{\beta} = S_\beta^\pm = 0$, $\sigma_S^2 = 1$.
Consequently, the 1-$\alpha$ Wald-type CIs are:
$$[S_L, S_U] = \left[\check S_\beta - z_{1-\alpha/2} \frac{\widehat\sigma_S}{\sqrt{n}}, \quad \check S_\beta + z_{1-\alpha/2} \frac{\widehat\sigma_S}{\sqrt{n}}\right],$$
where $z_{1-\alpha/2}$ is the $1-\alpha/2$ quantile of the standard normal distribution and $\widehat\sigma_S$ is consistent estimator of standard error $\sigma_S$ from Theorem \ref{thm:theorem1}.

\section{Simulation}

\subsection{Simulation Settings}

To evaluate the finite-sample performance of the proposed confidence intervals, we conducted simulation studies under both linear and logistic regression models.

\paragraph{Linear Model} 
We generated continuous outcomes from the model $Y = \beta X + \varepsilon$, where $X \sim \text{Bernoulli}(0.4)$. To assess robustness against distributional assumptions, we considered three error-generating mechanisms ($\varepsilon$): (1) Homoskedastic Normal: $\varepsilon \sim \mathcal{N}(0, \sigma^2 = 2)$, serving as a baseline; (2) Homoskedastic Gamma: $\varepsilon = (G-\mu_G)$ with $G \sim \mathrm{Gamma}(\text{shape} = 1.200, \text{rate} = 0.775)$ and $\mu_G = 1.549$, skewed errors generated from a centered Gamma distribution; and (3) Heteroskedastic Mixed Normal: $\varepsilon|X=x \sim \mathcal{N}(0, \sigma_x^2)$ with unequal group variances ($\sigma_0 = 1.111, \sigma_1 = 3.333$). For comparability across scenarios, errors were rescaled to maintain a marginal variance of $\text{Var}(\varepsilon) = \sigma^2=2$.
We varied five population effect sizes $S_\beta \in \{0,\,0.25,\,0.5,\,0.75,\,1\}$ by choosing $\beta$ to satisfy $S_\beta^2=\beta^\top \Sigma_\beta^{-1}\beta$ under the model-implied $\Sigma_\beta$ and varied the sample size $n \in \{50, 100, 150, 200, 300, 400\}$. 
Point estimation used estimators for $\widehat{S}_{\beta}$ (unsigned) and $\check{S}_{\beta}$ (signed) based on the F and t distributions (Section \ref{sec:asymp_dist_f}).
To benchmark performance against traditional metrics, we compared the signed RESI with Cohen's $d$ (both signed measures) and the unsigned RESI with Cohen's $f$ (both unsigned measures) computed under the same designs.

\paragraph{Logistic Models} 

Binary outcomes were generated via $\text{logit}\{\Pr(Y=1|X)\} = \eta + \beta X$, with $X \sim \text{Bernoulli}(0.5)$.
To vary group balance, we varied the intercept $\eta \in \{0, -1, -2\}$ to create scenarios where the outcome was ``balanced", ``semi-balanced", and ``unbalanced".
Target effect sizes were set to $S_{\beta} \in \{0, 0.1, 0.2, 0.3, 0.4\}$ by choosing $\beta$ accordingly so that $S_\beta^2=\beta^\top \Sigma_\beta^{-1}\beta$, with sample sizes $n$ ranging from 50 to 1500. 
Point estimation used $\widehat{S}_{\beta}$ (unsigned) and $\check{S}_{\beta}$ (signed) based on Chi-square and Z statistics.

\paragraph{Implementation and Metrics} 
For each scenario, we constructed 95\% CIs using the proposed methods. 
We evaluated the performance of intervals based on both model-based (parametric) covariance and robust sandwich covariance (HC3 for linear, HC0 for logistic) using 1,000 simulations. For the linear model, we use HC3 because it provides the most stable heteroscedasticity correction in finite samples, whereas for the logistic model, we use HC0 because other HC variants require weight adjustments that depend on $\beta$, which makes their form intractable in this setting. Performance metrics included bias and the coverage probability of the 95\% CIs.

\subsection{Simulation Results}

\begin{figure}[h!]
    \centering
    \subfigure[Unsigned Effect Size\label{fig:sim_lm_unsigned}]{
    \includegraphics[width=\linewidth]{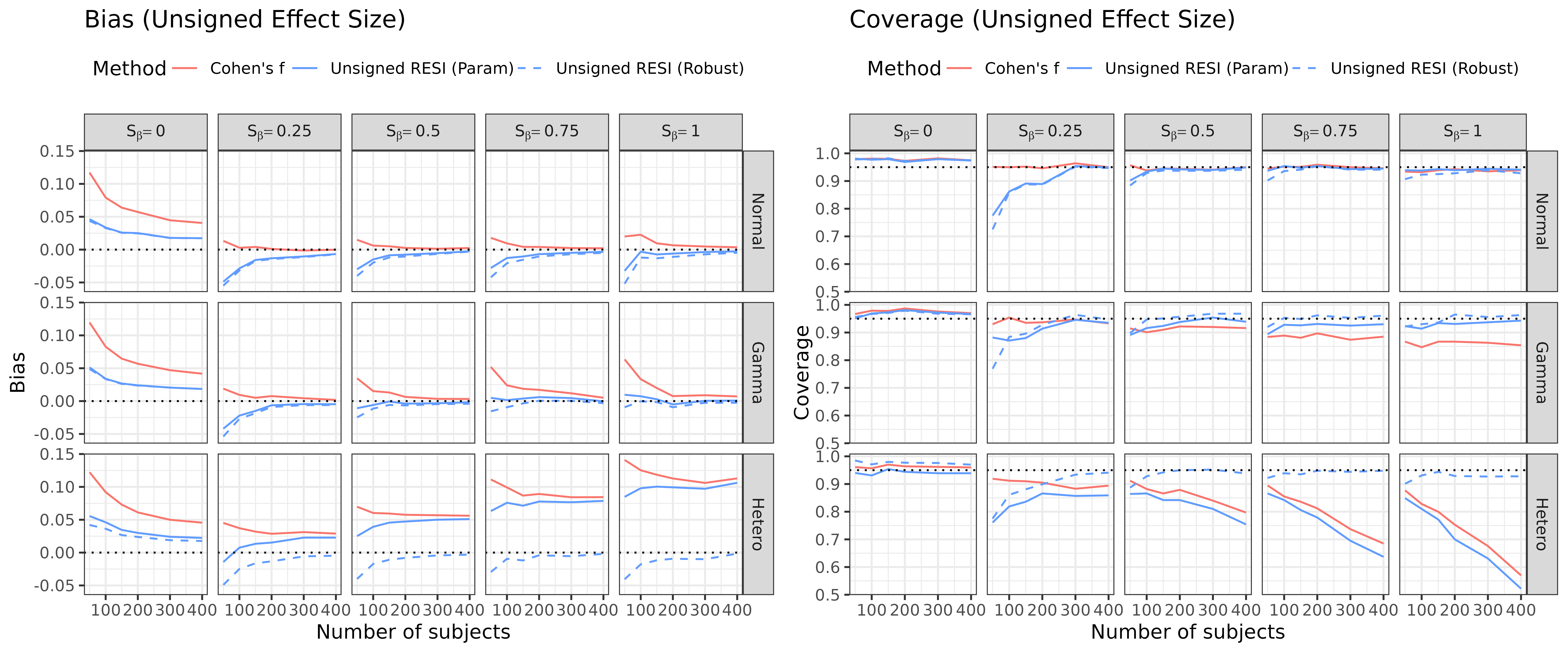}
    }
    \subfigure[Signed Effect Size\label{fig:sim_lm_signed}]{
    \includegraphics[width=\linewidth]{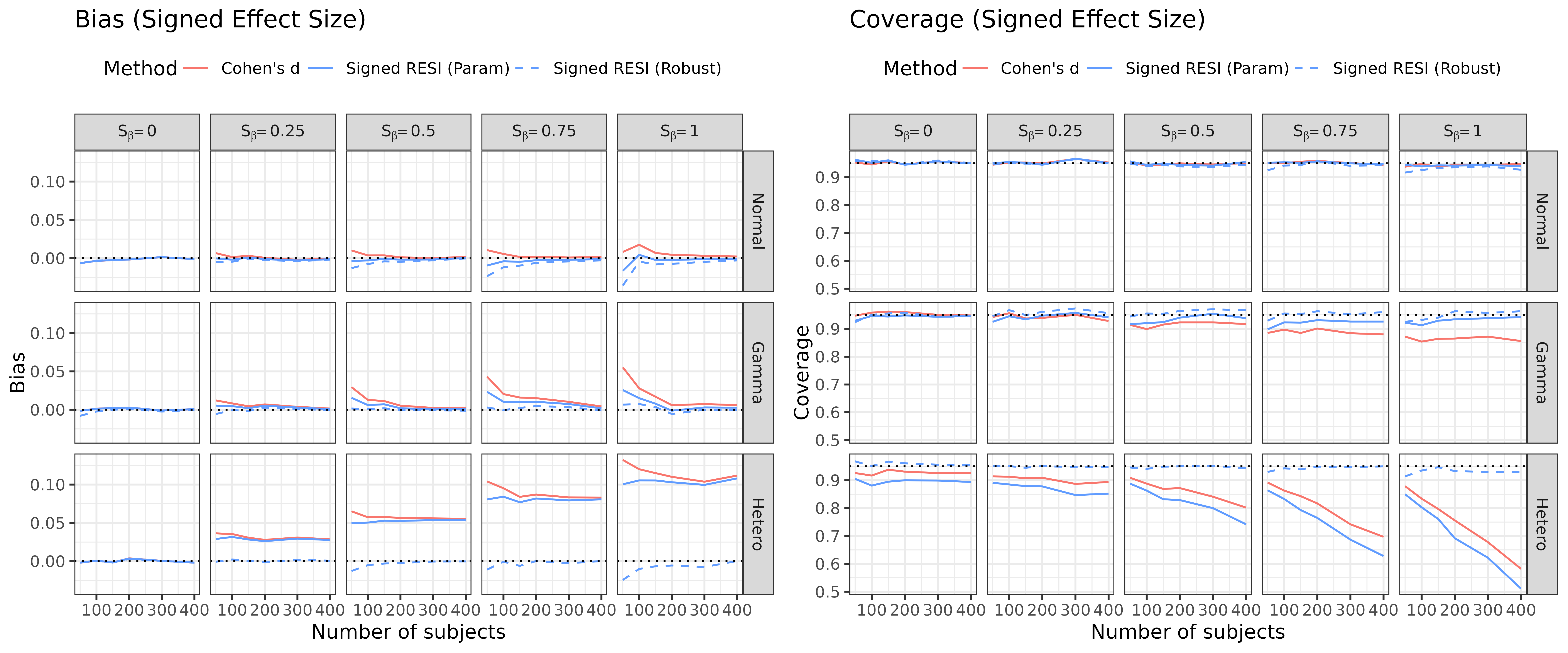}
    }
    \caption{Bias (left) and coverage (right) performance of the unsigned (a) and signed (b) RESI estimators under different true effect sizes (columns) and error distributions (rows).  ``Normal'' = homoskedastic, symmetric; ``Gamma'' = homoskedastic, asymmetric; ``Hetero'' = heteroskedastic. Curves show mean bias $\mathbb{E}[\widehat S_\beta - S_\beta]$ versus sample size; dashed line marks 0 for bias and 0.95 for coverage.}
\end{figure}

\paragraph{Linear model} 
Under the homoskedastic normal baseline, both Cohen's $f$ and the parametric and robust unsigned RESI estimators performed similarly with negligible bias and nominal coverage
(Figure \ref{fig:sim_lm_unsigned}).
RESI exhibited a slight positive bias near the boundary ($S_{\beta}=0$) due to its non-negative constraint, and Cohen's $f$ had a bias twice that of RESI.
While both RESI estimators performed well with the Gamma errors, the traditional standard indices (Cohen's $f$) showed substantial bias and under-coverage due to their normality assumption \citep{cohen_statistical_1988, hedges_statistical_1985}. 
Under heteroskedastic errors, the parametric RESI and Cohen's $f$ yielded decreasing coverage with increasing sample size, whereas the robust RESI (HC3) maintained valid inference.

The signed RESI and Cohen's $d$ showed markedly less bias with normal errors, and only the robust RESI was consistent under heteroskedasticity (Figure \ref{fig:sim_lm_signed}).
The confidence interval results were similar to the unsigned effect size.

\begin{figure}[h!]
    \centering
    \includegraphics[width=\linewidth]{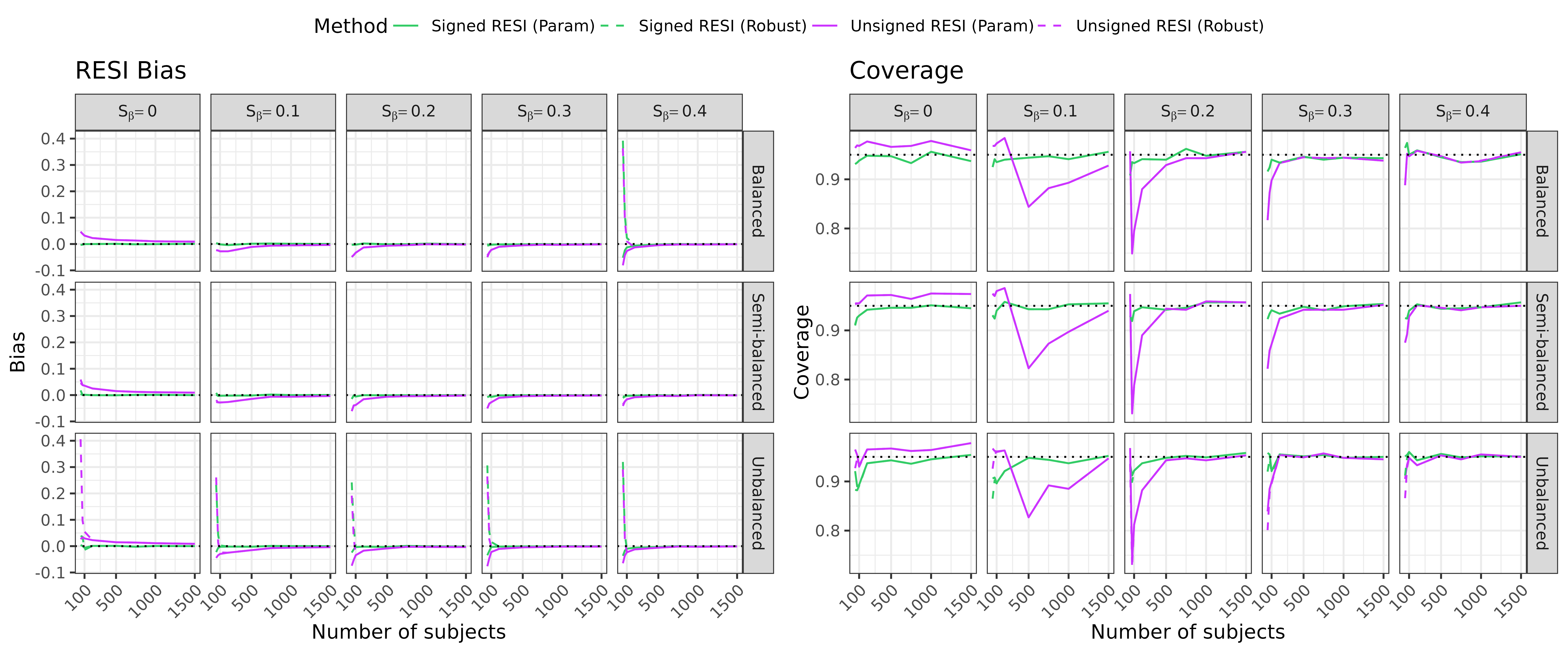}
    \caption{Bias (left) and coverage (right) of signed and unsigned RESI estimator by true effect sizes (columns) and intercepts (rows). ``Balanced'': $\eta=0$; ``Semi-balanced'': $\eta = -1$; ``Unbalanced'': $\eta = -2$. Curves show mean bias $\mathbb{E}[\widehat S_\beta - S_\beta]$ versus sample size; dashed line marks 0 for bias and 0.95 for coverage. }
    \label{fig:logistic}
\end{figure}

\paragraph{Logistic model}
With the logistic regression model, the parametric approach yielded less biased estimates than the robust estimator in small samples ($n < 100$; Figure \ref{fig:logistic}). This reflects the finite-sample instability of sandwich estimators in logistic regression with rare events.
Regarding estimator types, the signed RESI maintained consistent nominal coverage across most scenarios for sample sizes above 500.
The unsigned RESI showed noticeable coverage dips for small effect sizes ($S_{\beta} \le 0.2$) in smaller samples, a behavior attributable to the boundary properties of non-negative estimators; however, coverage improved with larger sample sizes or effect sizes.

\section{Characterizing Developmental Differences in ASD}

We use the bootstrap RESI CI \citep{kang_accurate_2023} and proposed CI to analyze nonlinear developmental differences between TD and ASD in childhood depression in 245 youth from the Pubertal Development Cohort (PDC).
Second, we analyze sex differences in the total problems score of the Child Behavioral Checklist (CBCL) \citep{achenbach_manual_2001} in 20,094 children and adolescents from the Simons Foundation Powering Autism Research cohort (SPARK) \citep{feliciano_spark_2018}.
We use Cohen's recommended values for small (RESI=0.1), medium (RESI=0.25), and large (RESI=0.4) effects \citep{cohen_statistical_1988,vandekar_robust_2020}.

\subsection{Diagnostic Developmental Differences in Depression}

The VUMC cohort was collected as part of a longitudinal study, examining stress and arousal across pubertal development in autism, recruited from a broad community sample in the southern US.
Eligibility criteria required an IQ $\ge 70$ and enrollment between 10.0 and 13.9 years of age.
Data were collected as part of a 4-year longitudinal study on pubertal development and stress, with diagnostic procedures completed at Year 1 (Y1) and physical examinations and psychological assessments conducted annually \citep{corbett_trajectory_2024}.
Here, we analyze a cross-sectional subset of Y1 data, which included 245 youth (140 ASD, 105 TD; Table \ref{tab:Y1_data}).
Details of sample ascertainment are described in prior work \citep{corbett_trajectory_2024}.

\begin{table}[h!]
\centering
\caption{PDC baseline characteristics by group in year 1.}
\begin{tabular}{lcc}
\toprule
 & TD ($n$=105) & ASD ($n$=140) \\
\midrule

\multicolumn{3}{l}{\textbf{Age (years)}} \\
\quad Mean (SD)                & 11.7 (1.22)            & 11.4 (1.03) \\
\quad Median [Min, Max]        & 11.7 [10.0, 13.9]       & 11.3 [10.0, 13.8] \\
\midrule

\multicolumn{3}{l}{\textbf{Sex}} \\
\quad Male                     & 59 (56.2\%)            & 104 (74.3\%) \\
\quad Female                   & 46 (43.8\%)            & 36 (25.7\%) \\
\midrule


\multicolumn{3}{l}{\textbf{Race}} \\
\quad White                        & 90 (85.7\%)            & 114 (81.4\%) \\
\quad Black                        & 2 (1.9\%)              & 17 (12.1\%) \\
\quad  Asian or Pacific Islander                        & 0 (0\%)                & 1 (0.7\%) \\
\quad Multiracial                        & 13 (12.4\%)            & 8 (5.7\%) \\
\midrule





\multicolumn{3}{l}{\textbf{CDI Total T-score}} \\
\quad Mean (SD)                & 51.1 (8.63)            & 58.7 (12.4) \\
\quad Median [Min, Max]        & 49 [40, 77]            & 57 [41, 90] \\
\quad Missing                  & 3 (2.9\%)              & 5 (3.6\%) \\
\bottomrule
\end{tabular}
\label{tab:Y1_data}
\end{table}

We examined diagnostic differences in early developmental trajectories of depression, measured by the Children's Depression Inventory (CDI) Total T-score \citep{kovacs_cdi_1985}.
We fitted a linear regression model with sex, diagnosis, age (modeled via natural cubic splines, df=3), and a diagnosis-by-age interaction.
Test statistics were computed using the HC3 robust covariance estimator, and ANOVA was performed using Type 2 sum of squares, so that the main effects are tested without the interaction term in the model.
Unsigned RESI CIs were computed for the ANOVA table, and signed RESI CIs were computed for the individual coefficients.

\begin{figure}[h!]
    \centering
    \includegraphics[width=0.8\linewidth]{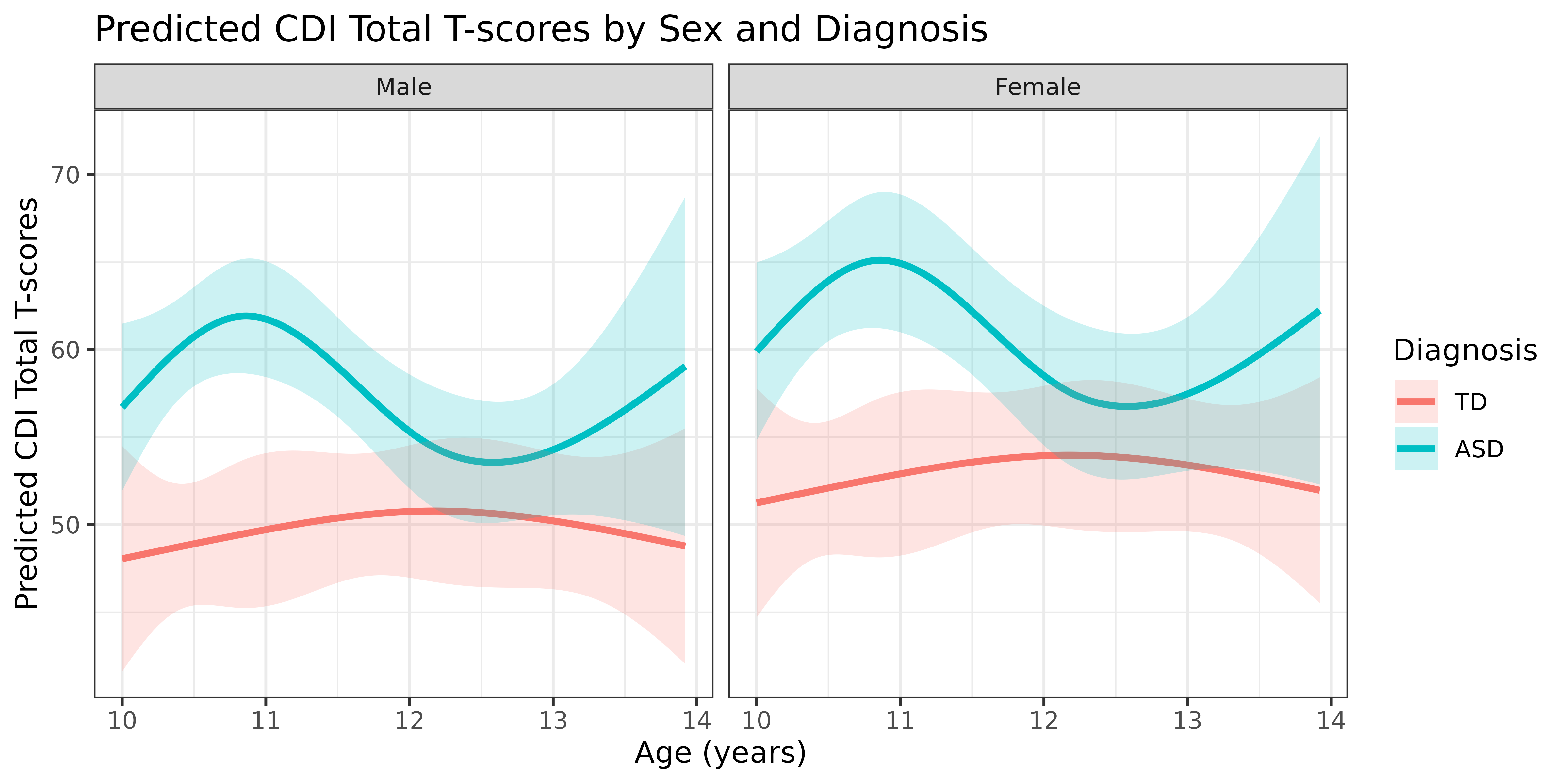}
    \caption{The model-estimated CDI Total score trajectories for ASD and TD across age and sex, with shaded bands indicating 95\% confidence intervals. Lines represent the predicted mean scores at each age, while the ribbons reflect uncertainty around those estimates. This visualization illustrates how children’s depression severity changes with age and whether the developmental patterns differ between ASD/TD and males/females.}
    \label{fig:dx_tscore}
\end{figure}

Diagnosis exhibited a large effect size ($\text{RESI}=0.39,\ \text{CI}=[0.27, 0.52],\ p<0.001$; Table \ref{tab:resi_anova_autism}), indicating a difference in depressive symptoms between ASD and TD youth. The ASD group consistently showed higher levels of depression across all ages and sexes (Figure \ref{fig:dx_tscore}). The confidence interval was wholly above a medium effect size index (0.25), suggesting with high confidence that diagnostic differences have a large effect. 
While, sex is statistically significant ($p=0.044$), its magnitude (RESI = 0.11, $\text{CI}=[0.00, 0.22]$) indicates a modest effect relative to diagnostic difference, and the nonlinear age effect (RESI = 0.00, $\text{CI}=[0.00, 0.11]$, $p=0.473$) remains small within this 10–14 year range, with the entire CI almost lying below the conventional threshold for a small effect (0.1).
The diagnosis-by-age interaction shows an effect similar in magnitude to the sex differences (RESI = 0.11, \text{CI}=[0.00, 0.22]) yet with a nonsignificant p-value ($p=0.111$), indicating only limited age-dependent variation in the magnitude of ASD–TD differences. This contrast illustrates that reliance on p-values alone can be misleading, as effects with comparable magnitude and uncertainty may be classified differently as ``significant'' or ``nonsignificant''. Altogether, the RESI values demonstrate that diagnosis is a dominant contributor to variation in CDI total T-scores, with sex, age, and their interaction explaining less variation. 

The asymptotic CIs were narrower than those derived from the bootstrap in Tables \ref{tab:resi_anova_autism} and \ref{tab:resi_coef_autism}. Given the reliable coverage observed in our simulations, this reduced interval width suggests the asymptotic approach offers greater precision without sacrificing statistical validity.
The asymptotic approach was approximately 53 times faster than the bootstrap (0.41 vs. 21.73 seconds).

\begin{table}[h!]
\centering
\caption{Type II ANOVA table of unsigned RESI estimates and two confidence intervals. 
LCI: lower confidence interval; UCI: upper confidence interval; Asymp: asymptotic CI; Boot: bootstrap CI. ns denotes the spline terms.}
\resizebox{\textwidth}{!}{%
\begin{tabular}{lcccccccc}
\toprule
Term & DF & RESI & SE & LCI Asymp & UCI Asymp & LCI Boot & UCI Boot & $p$-value \\
\midrule

diagnosis            & 1 & 0.39 & 0.06 & 0.27 & 0.52 & 0.26 & 0.54 & $<$0.001 \\

ns(age, 3)           & 3 & 0.00 & 0.07 & 0.00 & 0.11 & 0.00 & 0.28 & 0.473 \\

sex                  & 1 & 0.11 & 0.07 & 0.00 & 0.22 & 0.00 & 0.24 & 0.044 \\

diagnosis:ns(age, 3) & 3 & 0.11 & 0.07 & 0.00 & 0.22 & 0.00 & 0.29 & 0.111 \\

\bottomrule
\end{tabular}}
\label{tab:resi_anova_autism}
\end{table}

\begin{table}[h!]
\centering
\caption{Coefficient estimates, signed RESI estimates, and two confidence intervals. 
LCI: lower confidence interval; UCI: upper confidence interval; Asymp: asymptotic CI; Boot: bootstrap CI. ns denotes the spline terms.}
\resizebox{\textwidth}{!}{
\begin{tabular}{lcccccccc}
\toprule
Term & Estimate & RESI & SE & LCI Asymp & UCI Asymp & LCI Boot & UCI Boot & $p$-value \\
\midrule

(Intercept)              
    & 48.05 & 1.29 & 0.07 & 1.16 & 1.42 & 0.91 & 1.86 & $<$0.001 \\

diagnosis               
    & 8.65  & 0.15 & 0.07 & 0.02 & 0.28 & 0.02 & 0.30 & 0.019 \\

ns(age, 3)1              
    & 3.00  & 0.05 & 0.06 & -0.07 & 0.18 & -0.08 & 0.17 & 0.406 \\

ns(age, 3)2              
    & 3.18  & 0.04 & 0.06 & -0.09 & 0.16 & -0.08 & 0.18 & 0.582 \\

ns(age, 3)3              
    & -0.18 & -0.00 & 0.07 & -0.13 & 0.12 & -0.15 & 0.10 & 0.953 \\

sex                
    & 3.19  & 0.13 & 0.07 & 0.00 & 0.26 & 0.02 & 0.25 & 0.044 \\

diagnosis:ns(age, 3)1   
    & -13.40 & -0.14 & 0.07 & -0.27 & -0.01 & -0.27 & -0.02 & 0.029 \\

diagnosis:ns(age, 3)2   
    & 3.06   & 0.02 & 0.06 & -0.11 & 0.15 & -0.11 & 0.14 & 0.758 \\

diagnosis:ns(age, 3)3   
    & -2.26  & -0.02 & 0.07 & -0.15 & 0.11 & -0.19 & 0.09 & 0.753 \\

\bottomrule
\end{tabular}}
\label{tab:resi_coef_autism}
\end{table}

\subsection{Sex Effects on CBCL Total Problems T-score}

We applied the RESI framework to data from the SPARK cohort, a large-scale study of autism spectrum disorder \citep{feliciano_spark_2018}.
The SPARK sample comprises 20,094 autistic individuals (5,081 females, 15,013 males; Table \ref{tab:table1_spark}) ages 6 to 17 who completed the CBCL.
The CBCL is a widely administered parent/caregiver questionnaire to identify internalizing (e.g., anxiety, withdrawal) and externalizing (e.g., aggression, rule-breaking) behavioral problems \citep{achenbach_manual_2001}. 
We focused our analysis on the dichotomized Total Problems T-scores using a cutoff of 65, which is commonly used in clinical practice. T-scores $\ge$ 65 indicate clinically significant behavioral problems, while scores $<$ 65 are considered within the normative or borderline range, thereby providing a clinically interpretable distinction between affected and non-affected individuals \citep{achenbach_manual_2001}.
To characterize developmental trajectories of behavioral severity, we modeled dichotomized T-scores as a function of sex, age (modeled using natural cubic splines, df = 3), and their interaction, using a logistic regression.
As above, the HC3 robust covariance estimator and Type 2 sum of squares ANOVA were used for the analyses.

\begin{table}[h!]
\centering
\caption{SPARK dataset sample characteristics.}  
\label{tab:table1_spark}
\begin{tabular}{lcc}
\toprule
 & \textbf{Female ($n$=5,081)} & \textbf{Male ($n$=15,013)} \\
\midrule

\textbf{Child age (years)} & & \\
\quad Mean (SD)            & 10.5 (3.48)  & 10.4 (3.38) \\
\quad Median [Min, Max]    & 10.0 [6.00, 17.0] & 10.0 [6.00, 17.0] \\
\midrule

\textbf{CBCL Internalizing Problems T-score} & & \\
\quad Mean (SD)            & 64.1 (11.5)  & 62.7 (10.1) \\
\quad Median [Min, Max]    & 65.0 [33.0, 100] & 65.0 [34.0, 100] \\
\midrule

\textbf{CBCL Externalizing Problems T-score} & & \\
\quad Mean (SD)            & 59.8 (10.9)  & 58.6 (11.0) \\
\quad Median [Min, Max]    & 60.0 [34.0, 98.0] & 59.0 [33.0, 99.0] \\
\midrule

\textbf{CBCL Total Problems T-score} & & \\
\quad Mean (SD)            & 66.5 (8.94)  & 64.8 (9.30) \\
\quad Median [Min, Max]    & 68.0 [24.0, 99.0] & 65.0 [24.0, 95.0] \\
\bottomrule
\end{tabular}
\end{table}

\begin{figure}[h!]
    \centering
    \includegraphics[width=0.6\linewidth]{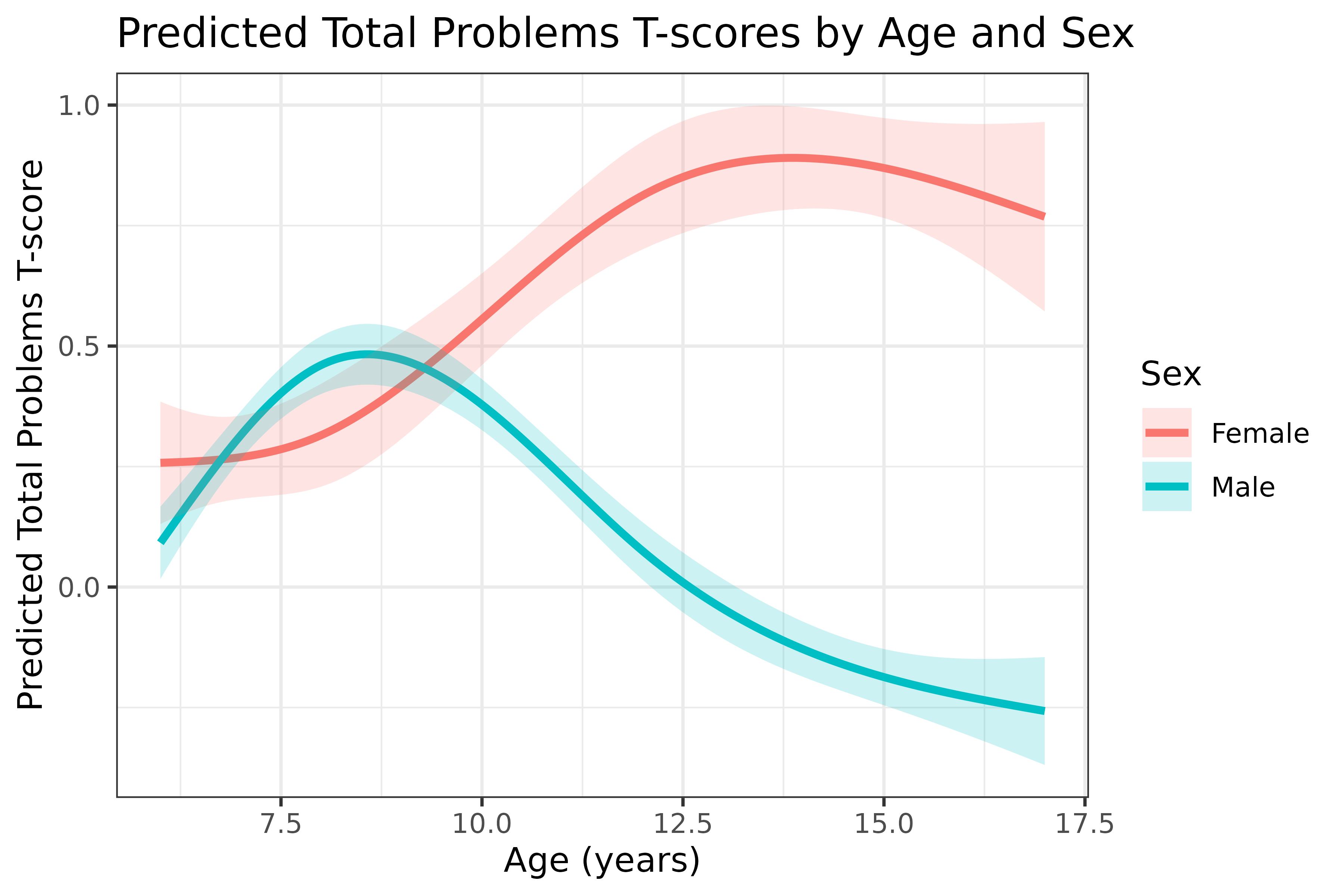}
    \caption{The model-estimated dichotomized Total Problems T-score trajectories for males and females across age, with shaded bands indicating 95\% confidence intervals. Lines represent the predicted mean scores at each age, while the ribbons reflect uncertainty around those estimates. This visualization illustrates how behavioral problem severity changes with age and whether the developmental patterns differ between males and females.}
    \label{fig:sex_tscore_glm}
\end{figure}

The ANOVA results (Table \ref{tab:resi_anova_spark_glm}) and coefficient estimates (Table \ref{tab:resi_coef_spark_glm}) with RESI using robust covariance reveal distinct sex-specific developmental patterns. 
From Table \ref{tab:resi_anova_spark_glm}, the sex effect corresponds to a RESI of 0.08 with a confidence interval of [0.07, 0.10], indicating a small difference across development in the overall burden of emotional and behavioral difficulties. Boys show higher scores in early childhood that decline with age, whereas girls’ scores increase with age and surpass boys in mid-childhood (Figure \ref{fig:sex_tscore_glm}).
The age effect (RESI = 0.07, \text{CI} = [0.06, 0.09], $p<0.001$) captures the nonlinear developmental pattern in symptom burden across both sexes. The interaction between sex and age has a comparable effect (RESI = 0.09, \text{CI} = [0.08, 0.11], $p<0.001$), indicating a small sex difference in the developmental pattern.
Although both sexes exhibit an overall inverted-U developmental pattern, boys show an earlier peak followed by a sharper decline in symptom burden during adolescence, whereas girls peak later and display a more gradual decrease.
Importantly, the confidence intervals for the sex, age, and sex-by-age interaction effects all lie mostly below a ``small" effect size (0.1), indicating that these developmental differences, while significant in a large sample, are modest in magnitude.
\\

Across all terms in Table \ref{tab:resi_anova_spark_glm} and \ref{tab:resi_coef_spark_glm}, the asymptotic and bootstrap CIs are nearly indistinguishable. 
This consistency is theoretically expected with a dataset of more than 20,000 observations, where the sampling distribution of the RESI estimates is well-approximated, and the bootstrap and asymptotic methods effectively converge to the same underlying distribution. However, the computational cost differs substantially: the asymptotic method completes in 11.04 seconds compared to 156.17 seconds for the bootstrap, making it roughly fifteen times faster. 

\begin{table}[h!]
\centering
\caption{Type II ANOVA table of unsigned RESI estimates and two confidence intervals. 
LCI: lower confidence interval; UCI: upper confidence interval; Asymp: asymptotic CI; Boot: bootstrap CI. ns denotes the spline terms.}
\resizebox{\textwidth}{!}{
\begin{tabular}{lcccccccc}
\toprule
Term & Df & RESI & SE & LCI Asymp & UCI Asymp & LCI Boot & UCI Boot & $p$-value \\
\midrule

sex                                 & 1 & 0.08 & 0.01 & 0.07 & 0.10 & 0.07 & 0.10 & $<0.001$ \\

ns(age, 3)         & 3 & 0.07 & 0.01 & 0.06 & 0.09 & 0.06 & 0.08 & $<0.001$ \\

sex:ns(age, 3)     & 3 & 0.09 & 0.01 & 0.08 & 0.11 & 0.08 & 0.11 & $<0.001$ \\

\bottomrule
\end{tabular}}
\label{tab:resi_anova_spark_glm}
\end{table}

\begin{table}[h!]
\centering
\caption{Coefficient estimates, signed RESI estimates, and two confidence intervals. 
LCI: lower confidence interval; UCI: upper confidence interval; Asymp: asymptotic CI; Boot: bootstrap CI. ns denotes the spline terms.}
\resizebox{\textwidth}{!}{
\begin{tabular}{lcccccccc}
\toprule
Term & Estimate & RESI & Std.Error & LCI Asymp & UCI Asymp & LCI Boot & UCI Boot & p-value \\
\midrule

(Intercept)                     
    & 0.26 & 0.03 & 0.01 & 0.01 & 0.04 & 0.01 & 0.04 & 0.0001 \\

sex                          
    & -0.17 & -0.02 & 0.01 & -0.03 & -0.00 & -0.03 & -0.00 & 0.027 \\

ns(age, 3)1      
    & 0.78 & 0.04 & 0.01 & 0.03 & 0.05 & 0.03 & 0.05 & $<0.001$ \\

ns(age, 3)2      
    & 0.62 & 0.03 & 0.01 & 0.01 & 0.04 & 0.01 & 0.04 & 0.0003 \\

ns(age, 3)3      
    & 0.50 & 0.03 & 0.01 & 0.02 & 0.05 & 0.02 & 0.05 & $<0.001$ \\

sex:ns(age, 3)1 
    & -1.18 & -0.05 & 0.01 & -0.07 & -0.04 & -0.07 & -0.04 & $<0.001$ \\

sex:ns(age, 3)2 
    & -0.40 & -0.01 & 0.01 & -0.03 & -0.00 & -0.03 & -0.00 & 0.045 \\

sex:ns(age, 3)3 
    & -1.19 & -0.07 & 0.01 & -0.08 & -0.06 & -0.08 & -0.06 & $<0.001$ \\

\bottomrule
\end{tabular}}
\label{tab:resi_coef_spark_glm}
\end{table}

As a sensitivity analysis, we additionally analyzed continuous T-scores using linear models, with results reported in Section \ref{sec:spark_lm_results}.

\newpage

\section{Discussion}


We developed an analytic method to construct CIs for the Robust Effect Size Index (RESI). Previously, RESI CI estimation required bootstrapping, which is computationally slow.
We derived the closed-form asymptotic distribution based on the estimating equation of the underlying models. This provides a fast and reliable alternative that works for a wide range of models, allowing researchers to report effect sizes with confidence intervals without the high computational cost.
Our analyses of developmental associations in ASD across two datasets with different sample sizes and model types reiterate the benefit of reporting the RESI and confidence intervals. 


In models that include nuisance parameters such as dispersion, simply treating these components as fixed can lead to underestimating uncertainty. To address this, we apply a joint estimation approach that accounts for the full set of model parameters and use the sandwich covariance estimator by default. This ensures that the estimated variability of RESI properly reflects uncertainty from both the effect of interest and other model components. When parametric variance is used, this approach leads to a simplified and accurate approximation of the RESI distribution, supporting valid inference and confidence intervals. 



The proposed RESI CI procedure confers three advantages to existing alternatives.
A significant advantage of the analytic approach is speed.
The new procedure is up to 50 times faster than bootstrapping, offering scalability critical for modern high-dimensional research.
Second, in the ASD developmental associations, the asymptotic confidence intervals for the unsigned RESI were narrower in the PDC sample of 245 participants than those derived from the bootstrap. Since our simulations confirm that the asymptotic method maintains nominal coverage, the narrower confidence intervals suggest it may be more precise than the bootstrap in this setting.
Lastly, our simulation results demonstrate why RESI is a more reliable metric than traditional indices like Cohen's $d$ and $f$: CIs of these conventional measures assume the data are normally distributed and that groups have equal variances. As shown in our simulations, when these assumptions fail (e.g., with skewed data or unbalanced groups), Cohen's $d$ and $f$ produce biased estimates and poor confidence interval coverage.
In contrast, RESI, when using with robust sandwich variance estimation, maintains valid inference even under model misspecification.
Due to the complex heterogeneity of outcomes in bio-behavioral medical research, these advantages make RESI a far safer choice for reporting effect sizes and confidence intervals for diverse datasets.







\backmatter


\section*{Acknowledgements}

This work was supported by the National Institutes of Health [R01MH123563 to Simon Vandekar] and the National Institute of Mental Health [MH111599 to Blythe Corbett].

Approved researchers can obtain the SPARK population dataset described in this study (\url{https://www.sfari.org/resource/spark/spark-phenotypic-data}) by applying at \url{https://base.sfari.org}.


\section*{Supplementary Materials}

\setcounter{table}{0}
\setcounter{figure}{0}

\renewcommand{\thetable}{S\arabic{table}}
\renewcommand{\thefigure}{S\arabic{figure}}
\renewcommand{\theequation}{S\arabic{equation}}
\renewcommand{\thesection}{S\arabic{section}} 

\subsection{Asymptotic Normality Conditions}\label{sec:regularity}

The following regularity conditions are required for the asymptotic normality of 
$\sqrt{n}(\hat{\theta} - \theta)$ \citep{van_der_vaart_asymptotic_2000, vandekar_robust_2019}:

\begin{enumerate}
    \item The function $\theta^{*} \mapsto \psi(\theta^{*}; W)$ is almost surely differentiable at $\theta$ (see Section \ref{sec:estimatingeq}).

    \item For every $\theta_1$ and $\theta_2$ in a neighborhood of $\theta$ and measurable function $m(W)$ such that $\mathbb{E}\ m(W)^2 < \infty$, 
    $$
        |\psi(\theta_1; W) - \psi(\theta_2; W)| \le m(W)\,\|\theta_1 - \theta_2\|.
    $$

    \item The function $\theta^{*} \mapsto \mathbb{E}\ \psi(\theta^{*}; W)$ admits a first-order Taylor expansion at $\theta$ with a non-singular derivative matrix $A(\theta)$.

    \item $\psi(\hat{\theta}, W) \ge \sup_{\theta^{*}} \psi(\theta^{*}, W) - o_p(n^{-1})$ and $\hat{\theta} \xrightarrow{p} \theta$.
\end{enumerate}

\subsection{Asymptotic Equivalence of Estimators}\label{sec:asy_equiv_dist}

\subsubsection{Estimator based on $T^2$ statistic}\label{sec:asymp_dist_t2}


Estimators $\widehat S_\beta$ \eqref{eq:shat_chisq} and $\widetilde S_\beta$ \eqref{eq:stilde} are asymptotically equal when $S_\beta>0$.
Note that
$$\sqrt{n}(\widehat S_\beta - S_\beta)
= \sqrt{n}(\widetilde S_\beta - S_\beta) + \sqrt{n}(\widehat S_\beta - \widetilde S_\beta).$$
Let 
$$
\widehat S_\beta = \left\{\max\left(0, \frac{T^2-m_1}{n}\right)\right\}^{1/2},
\qquad 
\widetilde S_\beta = \sqrt{\frac{T^2}{n}}.
$$
Under the alternative, we have $\frac{T^2}{n} \xrightarrow{p} S_\beta^2>0$, $T^2 = O_p(n)$, which implies $T^2 \xrightarrow{p} \infty$ and hence $\mathbb{P}(T^2 - m_1) \to 1$.
Decompose $\sqrt{n}(\widehat S_\beta - \widetilde S_\beta)$ by the truncation event, 
\begin{equation*}
\sqrt{n} (\widehat S_\beta - \widetilde S_\beta) = \boldsymbol{1}(T^2 > m_1)\left(\sqrt{T^2 - m_1}- \sqrt{T^2}\right) + \boldsymbol{1}(T^2 \le m_1)\left(0-\sqrt{T^2}\right)
\end{equation*}
The event $\{T^2 > m_1\}$, the first term yields $ - \frac{m_1}{\sqrt{T^2-m_1} + \sqrt{T^2}} = O_p(n^{-1/2})$. On the complement $\{T^2 \le m_1\}$, we have $\mid\sqrt{T^2} \mid \le \sqrt{m_1}$, hence $|0-\sqrt{T^2}| \le \sqrt{m_1} = o_p(1)$ because $\mathbb{P}(T^2 - m_1) \to 1$. Combining the two parts gives 
$\sqrt{n}(\widehat S_\beta - \widetilde S_\beta) = O_p(n^{-1/2})$, and 
$\sqrt{n}(\widetilde S_\beta - S_\beta) \xrightarrow{d} N(0,\sigma_S^2)$ by Theorem \ref{thm:theorem1}. Thus 
$\sqrt{n}(\widehat S_\beta - S_\beta) \xrightarrow{d} N(0, \sigma_S^2)$ by Slutsky’s theorem.

The convergence does not hold when $S_\beta=0$. Then $\sqrt{n}\widetilde S_\beta \xrightarrow{d} \chi_{m_1}$ converges to a central chi distribution on $m_1$ degrees of freedom,  and $\sqrt{n}\widehat S_\beta \xrightarrow{d} \sqrt{(\chi_{m_1}^2 - m_1)_+}$.
Their difference is non-degenerate
$$\sqrt{n}(\widehat S_\beta - \widetilde S_\beta) = \sqrt{(\chi_{m_1}^2-m_1)_+} - \sqrt{\chi_{m_1}^2} \not\xrightarrow{p} 0$$
Therefore, neither their asymptotic variance nor distribution is the same under the null.

\subsubsection{Estimator based on $Z$ statistic}\label{sec:asymp_dist_z}

Estimators $\sqrt{n}(\check S_\beta - S_\beta^{\pm})$ in \eqref{eq:scheck_z} and $\sqrt{n}(\widetilde S_\beta - S_\beta)$ in \eqref{eq:stilde} are asymptotically equal in distribution when $S_\beta>0 \ (m_1=1)$, where $S_\beta^{\pm} = sgn(\beta - \beta_0)S_\beta$.

Let 
$$\check S_\beta = \frac{Z}{\sqrt{n}},\quad \widetilde S_\beta = \sqrt{\frac{T^2}{n}},\ \text{with}\ Z = \frac{\hat\beta - \beta_0}{\hat{se}(\hat\beta)}, \quad T^2 = Z^2.$$

Under the alternative, we have $\sqrt{n}(\check S_\beta - S_\beta^{\pm}) \xrightarrow{d}N(0, \sigma_1^2)$ by asymptotic normality of $\hat\beta$ and Slutsky's theorem, and $\sqrt{n}(\widetilde S_\beta - S_\beta) \xrightarrow{d} N(0, \sigma_2^2)$ by Theorem \ref{thm:theorem1}. 
By definition, $\displaystyle\widetilde S_\beta = \sqrt{\frac{T^2}{n}} = \frac{|Z|}{\sqrt{n}} = |\check S_\beta|$ and $\displaystyle S_\beta = |S_\beta^{\pm}|$. Therefore, $$\sqrt{n}(\widetilde S_\beta - S_\beta) = \sqrt{n}\left(|\check S_\beta| - |S_\beta^{\pm}|\right).$$ 
Let $\widetilde S_\beta = g(\check S_\beta)$, where $g(x) = |x|$. When $S_\beta > 0$ ($S_\beta^{\pm} \not=0$), $g(x)$ is differentiable at $x_0 = S_\beta^{\pm}$ with derivative $g'(x_0) = sgn(S_\beta^{\pm}) \in \{-1, 1\}$.  By the delta method, we have 
$$\sqrt{n}(\widetilde S_\beta - S_\beta) = \sqrt{n}\{g(\check S_\beta) - g(S_\beta^{\pm})\} \xrightarrow{d} N(0, g'(S_\beta^{\pm})^2\sigma^2_1) = N(0,\sigma^2_1).$$
Comparing this with the given convergence, the uniqueness of weak limits implies $\sigma_1^2 = \sigma^2_2$.

The convergence does not hold when $S_\beta=0\ (S_\beta^\pm=0)$.  $\sqrt n\,\widetilde S_\beta =  Z \Rightarrow N(0,\sigma_1^2)$.
Then
$$
\sqrt n(\widetilde S_\beta - S_\beta)=\sqrt n\,\widetilde S_\beta = |Z|
\Rightarrow |N(0,\sigma_1^2)|,
$$
a folded normal distribution with variance $\sigma_1^2(1-2/\pi)$, which differs from $\sigma_1^2$.

\subsubsection{Estimators based on $F$- and $t$-statistics}\label{sec:asymp_dist_f}

Under normality, we have $F = T^2/m_1$ \citep{mantel_chi-square_1963}.
The F-statistic-based estimator is
$$\displaystyle\hat{S}_\beta^{F} 
= \left\{ \max \left[ 0, 
\frac{Fm_1(n - m - 2) - m_1(n - m)}{n (n - m)} \right] \right\}^{1/2}$$.
This has the same asymptotic distribution as $\sqrt{n}(\widetilde S_\beta-S_\beta)$ under the alternative by the same argument as Section \ref{sec:asymp_dist_t2}.

For $m_1=1$, the $t$-statistic-based estimator is 
$$\check{S}_\beta^t = 
\frac{t \sqrt{2}\Gamma\!\left(\frac{n - m}{2}\right)}
{\sqrt{n(n - m)}\,\Gamma\!\left(\frac{n - m - 1}{2}\right)}$$
where
$$t = \frac{\hat\beta - \beta_0}{\hat\sigma_\beta}.$$ 

This has the same asymptotic distribution as $\sqrt{n}(\widetilde S_\beta-S_\beta)$ under the alternative by the same argument as Section \ref{sec:asymp_dist_z}.



\subsection{Derivatives}\label{sec:derivatives}

The following provides all derivatives required for the computation with robust covariance.
\begin{flalign*}
\frac{\partial S_\beta}{\partial \theta} = \frac{1}{S_\beta} L^\top \Sigma_\beta^{-1}L \theta \in \mathbb{R}^{m}&&
\end{flalign*}
\vspace{-4em}

\begin{flalign*}
\frac{\partial S_\beta}{\partial \text{vec}(B_\theta)} = -\frac{1}{2S_\beta}  (L A_\theta^{-1} \otimes L A_\theta^{-1})^\top \operatorname{vec}\left( \Sigma_\beta^{-1} L
\theta \theta^\top L^\top \Sigma_\beta^{-1} \right)  \in \mathbb{R}^{m^2} &&
\end{flalign*}
\vspace{-4em}

\begin{flalign*}
\frac{\partial S_\beta}{\partial \text{vec}(A_\theta)} = \frac{1}{2S_\beta}  \left\{(L A_\theta^{-1}  \otimes L A_\theta^{-1} B_\theta A_\theta^{-1}) + (LA_\theta^{-1} B_\theta A_\theta^{-1} \otimes L A_\theta^{-1}) \right\}^\top \operatorname{vec}(\Sigma_\beta^{-1} L
\theta \theta^\top L^\top \Sigma_\beta^{-1})  
 \in \mathbb{R}^{m^2}&&
\end{flalign*}
\vspace{-4em}

\begin{flalign*}
\frac{\partial \text{vec}(A_\theta)}{\partial \theta} = \frac{-d\ \mathbb{E}\{\Psi'(\theta; Y)\}}{d\theta} \in \mathbb{R}^{m \times m^2} &&
\end{flalign*}
\vspace{-4em}

\begin{flalign*}
\frac{\partial \text{vec}(B_\theta)}{\partial \theta} = \frac{d\ \mathbb{E}\{\Psi(\theta;Y)\Psi^\top(\theta;Y)\}}{d\theta} \in \mathbb{R}^{m \times m^2}&&
\end{flalign*}

\noindent For the parametric covariance, the derivative of $A_\theta$ with respect to $\theta$ is:
\vspace{-2em}

\begin{flalign*}
\frac{\partial S_\beta}{\partial \text{vec}(A_\theta)} = \frac{1}{2S_\beta} \operatorname{vec}^\top\left(\Sigma_\beta^{-1} L
\theta \theta^\top L^\top \Sigma_\beta^{-1} \right)(L A_\theta^{-1} \otimes L A_\theta^{-1})&&
\end{flalign*}

\subsection{Proof of Theorem 1}\label{sec:proof_thm1}

We establish the asymptotic normality of $\widetilde S_\beta = S_\beta(\hat{\theta})$ by invoking Z-estimation theory to guarantee the asymptotic normality of $\hat \theta$ and use a first-order delta-method to get the distribution of $\widetilde S_\beta$.

Let objects be as defined in Theorem \ref{thm:theorem1}. The regularity conditions in Section \ref{sec:regularity} are sufficient for
$$
\sqrt{n}(\hat{\theta}-\theta) \xrightarrow{d} \mathcal{N}(0,\Sigma_\theta),
$$
by Theorem 5.21 and 5.23 from \citep{van_der_vaart_asymptotic_2000}.

The multivariate delta method is valid with Condition $\it{(A2)}$ in Theorem \ref{thm:theorem1}.
This completes the proof.
The variance estimator is consistent by the Continuous Mapping Theorem.

\subsection{Limit of the asymptotic variance}\label{sec:limit_variance}

Since the unsigned estimators involve an absolute-value transformation, their limiting distributions under the null are non-normal and lead to a discontinuity in the asymptotic variance at the boundary. To study the limiting behavior of the variance in a unified manner, we therefore focus on the signed estimator, for which standard asymptotic normality holds. 

Using the same notation as in Section \ref{sec:theorem}, we define the signed RESI $S_\beta^{\pm}(\theta) = \beta^\top \Sigma_\beta^{-1/2} = \theta^\top L^\top \left(L A^{-1}_\theta B_\theta A^{-1}_\theta L^\top\right)^{-1/2}$. Theorem~1 remains valid for the estimator $\check S_\beta^{Z}$, with the corresponding derivatives under the robust covariance given by
\begin{flalign*}
\frac{\partial S_\beta^{\pm}}{\partial \theta}  =L^\top \Sigma_\beta^{-1/2} &&
\end{flalign*}
\vspace{-4em}

\begin{flalign*}
\frac{\partial S_\beta^{\pm}}{\partial \text{vec}(B_\theta)}  = -\frac{1}{2}\theta^\top L^\top \Sigma_\beta ^{-3/2} (LA_\theta^{-1}\otimes L A_\theta^{-1})&&
\end{flalign*}
\vspace{-4em}

\begin{flalign*}
\frac{\partial S_\beta^{\pm}}{\partial \text{vec}(A_\theta)}  = \frac{1}{2}\theta^\top L^\top \Sigma_\beta ^{-3/2} \left\{(L A_\theta^{-1}  \otimes L A_\theta^{-1} B_\theta A_\theta^{-1}) + (LA_\theta^{-1} B_\theta A_\theta^{-1} \otimes L A_\theta^{-1}) \right\}&&
\end{flalign*}

The derivatives $\partial \mathrm{vec}(A_\theta)/\partial \theta$ and
$\partial \mathrm{vec}(B_\theta)/\partial \theta$ remain the same as those given
in Section~\ref{sec:derivatives}. For the model-based case, the corresponding
derivative with respect to $\mathrm{vec}(A_\theta)$ simplifies to
\begin{flalign*}
\frac{\partial S_\beta^{\pm}}{\partial \text{vec}(A_\theta)} = \frac{1}{2}\theta^\top L^\top \Sigma_\beta ^{-3/2} (L A_\theta^{-1} \otimes L A_\theta^{-1})&&
\end{flalign*}

Under the null hypothesis $H_0: \beta = L\theta = 0$, both $\frac{\partial S_\beta}{\partial \text{vec}(A_\theta)}$ and $\frac{\partial S_\beta}{\partial \text{vec}(B_\theta)}$ vanish. By Theorem \ref{thm:theorem1}, the asymptotic variance $\sigma^2_S$ under the robust covariance $\Sigma_\theta$ is therefore
\begin{equation*}
\sigma_S^2 = \left(\frac{d S_\beta}{d \theta} \right)^\top \Sigma_\theta \left( \frac{d S_\beta}{d \theta} \right) = \left(\frac{\partial S_\beta}{\partial \theta}\right)^\top \Sigma_\theta \left(\frac{\partial S_\beta}{\partial \theta}\right) = (L^\top \Sigma^{-1/2}_\beta)^\top \Sigma_\theta L^\top \Sigma^{-1/2}_\beta = 1
\end{equation*}

\subsection{Additional application results}\label{sec:spark_lm_results}

\begin{figure}[h!]
    \centering
    \includegraphics[width=0.5\linewidth]{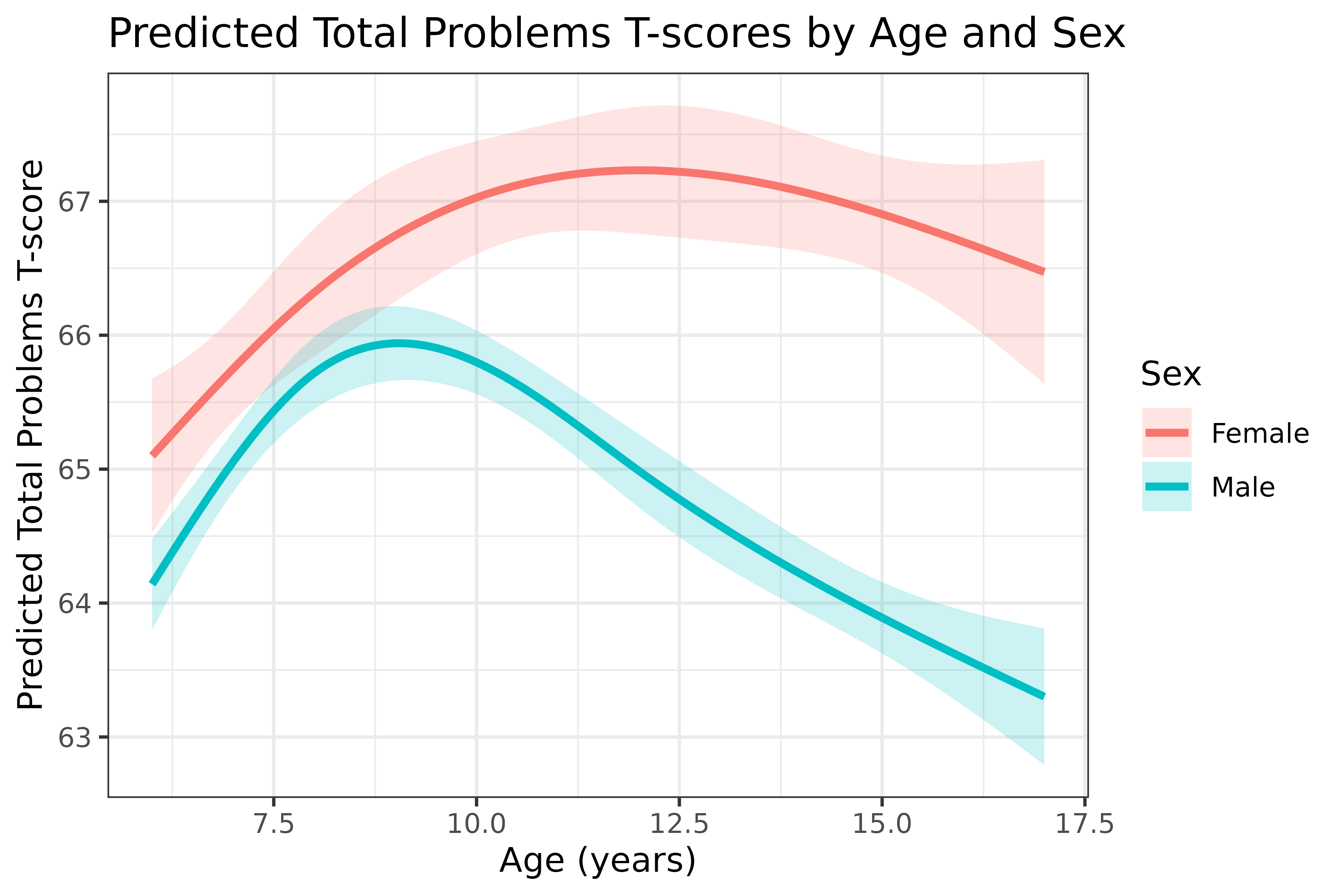}
    \caption{The model-estimated continuous Total Problems T-score trajectories for males and females across age, with shaded bands indicating 95\% confidence intervals. Lines represent the predicted mean scores at each age, while the ribbons reflect uncertainty around those estimates. This visualization illustrates how behavioral problem severity changes with age and whether the developmental patterns differ between males and females.}
    \label{fig:sex_tscore}
\end{figure}

\begin{table}[h!]
\centering
\caption{Type II ANOVA table of unsigned RESI estimates and two confidence intervals (linear model). 
LCI: lower confidence interval; UCI: upper confidence interval; Asymp: asymptotic CI; Boot: bootstrap CI. ns denotes the spline terms.}
\resizebox{\textwidth}{!}{
\begin{tabular}{lcccccccc}
\toprule
Term & Df & RESI & SE & LCI Asymp & UCI Asymp & LCI Boot & UCI Boot & $p$-value \\
\midrule

sex                                 & 1 & 0.08 & 0.01 & 0.07 & 0.10 & 0.07 & 0.10 & $<0.001$ \\

ns(age, 3)         & 3 & 0.07 & 0.01 & 0.06 & 0.08 & 0.06 & 0.09 & $<0.001$ \\

sex:ns(age, 3)     & 3 & 0.04 & 0.01 & 0.03 & 0.06 & 0.03 & 0.06 & $<0.001$ \\

\bottomrule
\end{tabular}}
\label{tab:resi_anova_spark}
\end{table}

\begin{table}[h!]
\centering
\caption{Coefficient estimates, signed RESI estimates, and two confidence intervals (linear model). 
LCI: lower confidence interval; UCI: upper confidence interval; Asymp: asymptotic CI; Boot: bootstrap CI. ns denotes the spline terms.}
\resizebox{\textwidth}{!}{
\begin{tabular}{lcccccccc}
\toprule
Term & Estimate & RESI & Std.Error & LCI Asymp & UCI Asymp & LCI Boot & UCI Boot & p-value \\
\midrule

(Intercept)                     
    & 65.10 & 1.53 & 0.01 & 1.51 & 1.55 & 1.46 & 1.61 & $<0.001$ \\

sex                          
    & -0.96 & -0.02 & 0.01 & -0.03 & -0.01 & -0.03 & -0.01 & 0.006 \\

ns(age, 3)1      
    & 1.91 & 0.02 & 0.01 & 0.01 & 0.04 & 0.01 & 0.04 & 0.001 \\

ns(age, 3)2      
    & 3.22 & 0.03 & 0.01 & 0.02 & 0.04 & 0.02 & 0.04 & $<0.001$ \\

ns(age, 3)3      
    & 0.43 & 0.01 & 0.01 & -0.01 & 0.02 & -0.01 & 0.02 & 0.298 \\

sex:ns(age, 3)1 
    & -2.17 & -0.02 & 0.01 & -0.04 & -0.01 & -0.04 & -0.01 & 0.001 \\

sex:ns(age, 3)2 
    & -1.46 & -0.01 & 0.01 & -0.03 & 0.00 & -0.02 & 0.00 & 0.103 \\

sex:ns(age, 3)3 
    & -2.63 & -0.04 & 0.01 & -0.05 & -0.02 & -0.05 & -0.02 & $<0.001$ \\

\bottomrule
\end{tabular}}
\label{tab:resi_coef_spark}
\end{table}



\bibliographystyle{biom}
\bibliography{MyLibrary1129}

@article{kang_accurate_2023,
	title = {Accurate {Confidence} and {Bayesian} {Interval} {Estimation} for {Non}-centrality {Parameters} and {Effect} {Size} {Indices}},
	volume = {88},
	issn = {1860-0980},
	url = {https://doi.org/10.1007/s11336-022-09899-x},
	doi = {10.1007/s11336-022-09899-x},
	language = {en},
	number = {1},
	urldate = {2024-03-21},
	journal = {Psychometrika},
	author = {Kang, Kaidi and Jones, Megan T. and Armstrong, Kristan and Avery, Suzanne and McHugo, Maureen and Heckers, Stephan and Vandekar, Simon},
	month = mar,
	year = {2023},
	pages = {253--273},
}

@article{kang_study_2024,
	title = {Study design features increase replicability in brain-wide association studies},
	volume = {636},
	copyright = {2024 The Author(s)},
	issn = {1476-4687},
	url = {https://www.nature.com/articles/s41586-024-08260-9},
	doi = {10.1038/s41586-024-08260-9},
	language = {en},
	number = {8043},
	urldate = {2025-01-30},
	journal = {Nature},
	author = {Kang, Kaidi and Seidlitz, Jakob and Bethlehem, Richard A. I. and Xiong, Jiangmei and Jones, Megan T. and Mehta, Kahini and Keller, Arielle S. and Tao, Ran and Randolph, Anita and Larsen, Bart and Tervo-Clemmens, Brenden and Feczko, Eric and Dominguez, Oscar Miranda and Nelson, Steven M. and Schildcrout, Jonathan and Fair, Damien A. and Satterthwaite, Theodore D. and Alexander-Bloch, Aaron and Vandekar, Simon},
	month = dec,
	year = {2024},
	note = {Publisher: Nature Publishing Group},
	pages = {719--727},
}

@book{boos_essential_2013,
	title = {Essential {Statistical} {Inference}: {Theory} and {Methods}},
	isbn = {978-1-4614-4818-1},
	shorttitle = {Essential {Statistical} {Inference}},
	language = {en},
	publisher = {Springer Science \& Business Media},
	author = {Boos, Dennis D. and Stefanski, L. A.},
	month = feb,
	year = {2013},
	note = {Google-Books-ID: 8VNDAAAAQBAJ},
}

@article{jones_resi_2025,
	title = {{RESI}: {An} {R} {Package} for {Robust} {Effect} {Sizes}},
	volume = {112},
	copyright = {Copyright (c) 2025 Megan Jones, Kaidi Kang, Simon Vandekar},
	issn = {1548-7660},
	shorttitle = {{RESI}},
	url = {https://doi.org/10.18637/jss.v112.i03},
	doi = {10.18637/jss.v112.i03},
	language = {en},
	urldate = {2025-05-12},
	journal = {Journal of Statistical Software},
	author = {Jones, Megan and Kang, Kaidi and Vandekar, Simon},
	month = mar,
	year = {2025},
	pages = {1--27},
}

@article{kafadar_editorial_2021,
	title = {Editorial: {Statistical} {Significance}, ?-{Values}, and {Replicability}},
	volume = {15},
	issn = {1932-6157},
	shorttitle = {Editorial},
	url = {https://www.jstor.org/stable/27210060},
	number = {3},
	urldate = {2025-08-27},
	journal = {The Annals of Applied Statistics},
	author = {Kafadar, Karen},
	year = {2021},
	note = {Publisher: Institute of Mathematical Statistics},
	pages = {1081--1083},
}

@article{mantel_chi-square_1963,
	title = {Chi-{Square} {Tests} with {One} {Degree} of {Freedom}; {Extensions} of the {Mantel}-{Haenszel} {Procedure}},
	volume = {58},
	issn = {0162-1459},
	url = {https://doi.org/10.1080/01621459.1963.10500879},
	doi = {10.1080/01621459.1963.10500879},
	number = {303},
	urldate = {2025-10-04},
	journal = {Journal of the American Statistical Association},
	author = {Mantel, Nathan},
	month = sep,
	year = {1963},
	note = {Publisher: ASA Website
\_eprint: https://doi.org/10.1080/01621459.1963.10500879},
	pages = {690--700},
}

@article{vandekar_robust_2020,
	title = {A {Robust} {Effect} {Size} {Index}},
	volume = {85},
	issn = {0033-3123, 1860-0980},
	url = {https://www.cambridge.org/core/journals/psychometrika/article/robust-effect-size-index/BDCC80E518297D91B344CE2D9291B2BC},
	doi = {10.1007/s11336-020-09698-2},
	language = {en},
	number = {1},
	urldate = {2025-11-07},
	journal = {Psychometrika},
	author = {Vandekar, Simon and Tao, Ran and Blume, Jeffrey},
	month = mar,
	year = {2020},
	pages = {232--246},
}

@article{liang_longitudinal_1986,
	title = {Longitudinal data analysis using generalized linear models},
	volume = {73},
	issn = {0006-3444},
	url = {https://doi.org/10.1093/biomet/73.1.13},
	doi = {10.1093/biomet/73.1.13},
	number = {1},
	urldate = {2025-11-12},
	journal = {Biometrika},
	author = {LIANG, KUNG-YEE and ZEGER, SCOTT L.},
	month = apr,
	year = {1986},
	pages = {13--22},
}

@article{corbett_trajectory_2024,
	title = {Trajectory of depressive symptoms over adolescence in autistic and neurotypical youth},
	volume = {15},
	issn = {2040-2392},
	url = {https://doi.org/10.1186/s13229-024-00600-w},
	doi = {10.1186/s13229-024-00600-w},
	language = {en},
	number = {1},
	urldate = {2025-11-19},
	journal = {Molecular Autism},
	author = {Corbett, Blythe A. and Muscatello, Rachael A. and McGonigle, Trey and Vandekar, Simon and Burroughs, Christina and Sparks, Sloane},
	month = may,
	year = {2024},
	pages = {18},
}

@article{olejnik_generalized_2003,
	title = {Generalized {Eta} and {Omega} {Squared} {Statistics}: {Measures} of {Effect} {Size} for {Some} {Common} {Research} {Designs}},
	volume = {8},
	issn = {1939-1463},
	shorttitle = {Generalized {Eta} and {Omega} {Squared} {Statistics}},
	doi = {10.1037/1082-989X.8.4.434},
	number = {4},
	journal = {Psychological Methods},
	author = {Olejnik, Stephen and Algina, James},
	year = {2003},
	note = {Place: US
Publisher: American Psychological Association},
	pages = {434--447},
}

@article{lord_autism_2020,
	title = {Autism spectrum disorder},
	volume = {6},
	copyright = {2020 Springer Nature Limited},
	issn = {2056-676X},
	url = {https://www.nature.com/articles/s41572-019-0138-4},
	doi = {10.1038/s41572-019-0138-4},
	language = {en},
	number = {1},
	urldate = {2025-11-26},
	journal = {Nature Reviews Disease Primers},
	author = {Lord, Catherine and Brugha, Traolach S. and Charman, Tony and Cusack, James and Dumas, Guillaume and Frazier, Thomas and Jones, Emily J. H. and Jones, Rebecca M. and Pickles, Andrew and State, Matthew W. and Taylor, Julie Lounds and Veenstra-VanderWeele, Jeremy},
	month = jan,
	year = {2020},
	note = {Publisher: Nature Publishing Group},
	pages = {5},
}

@article{cumming_new_2014,
	title = {The {New} {Statistics}: {Why} and {How}},
	volume = {25},
	issn = {0956-7976},
	shorttitle = {The {New} {Statistics}},
	url = {https://doi.org/10.1177/0956797613504966},
	doi = {10.1177/0956797613504966},
	language = {EN},
	number = {1},
	urldate = {2025-11-26},
	journal = {Psychological Science},
	author = {Cumming, Geoff},
	month = jan,
	year = {2014},
	note = {Publisher: SAGE Publications Inc},
	pages = {7--29},
}

@article{lakens_equivalence_2017,
	title = {Equivalence {Tests}: {A} {Practical} {Primer} for t {Tests}, {Correlations}, and {Meta}-{Analyses}},
	volume = {8},
	issn = {1948-5506},
	shorttitle = {Equivalence {Tests}},
	url = {https://doi.org/10.1177/1948550617697177},
	doi = {10.1177/1948550617697177},
	language = {EN},
	number = {4},
	urldate = {2025-11-26},
	journal = {Social Psychological and Personality Science},
	author = {Lakens, Daniël},
	month = may,
	year = {2017},
	note = {Publisher: SAGE Publications Inc},
	pages = {355--362},
}

@article{gotham_standardizing_2009,
	title = {Standardizing {ADOS} {Scores} for a {Measure} of {Severity} in {Autism} {Spectrum} {Disorders}},
	volume = {39},
	issn = {1573-3432},
	url = {https://doi.org/10.1007/s10803-008-0674-3},
	doi = {10.1007/s10803-008-0674-3},
	language = {en},
	number = {5},
	urldate = {2025-11-26},
	journal = {Journal of Autism and Developmental Disorders},
	author = {Gotham, Katherine and Pickles, Andrew and Lord, Catherine},
	month = may,
	year = {2009},
	pages = {693--705},
}

@article{happe_time_2006,
	title = {Time to give up on a single explanation for autism},
	volume = {9},
	copyright = {2006 Springer Nature America, Inc.},
	issn = {1546-1726},
	url = {https://www.nature.com/articles/nn1770},
	doi = {10.1038/nn1770},
	language = {en},
	number = {10},
	urldate = {2025-11-26},
	journal = {Nature Neuroscience},
	author = {Happé, Francesca and Ronald, Angelica and Plomin, Robert},
	month = oct,
	year = {2006},
	note = {Publisher: Nature Publishing Group},
	pages = {1218--1220},
}

@article{white_heteroskedasticity-consistent_1980,
	title = {A heteroskedasticity-consistent covariance matrix estimator and a direct test for heteroskedasticity},
	journal = {Econometrica: Journal of the Econometric Society},
	author = {White, Halbert},
	year = {1980},
	pages = {817--838},
}

@article{huber_robust_1964,
	title = {Robust {Estimation} of a {Location} {Parameter}},
	volume = {35},
	issn = {0003-4851, 2168-8990},
	url = {https://projecteuclid.org/euclid.aoms/1177703732},
	doi = {10.1214/aoms/1177703732},
	language = {EN},
	number = {1},
	urldate = {2018-07-24},
	journal = {The Annals of Mathematical Statistics},
	author = {Huber, Peter J.},
	month = mar,
	year = {1964},
	mrnumber = {MR161415},
	zmnumber = {0136.39805},
	pages = {73--101},
}

@book{van_der_vaart_asymptotic_2000,
	title = {Asymptotic statistics},
	volume = {3},
	publisher = {Cambridge university press},
	author = {Van der Vaart, Aad W.},
	year = {2000},
}

@book{hedges_statistical_1985,
	address = {London, UK},
	title = {Statistical {Methods} for {Meta}-{Analysis}},
	isbn = {978-0-08-057065-5},
	url = {https://linkinghub.elsevier.com/retrieve/pii/C20090033960},
	language = {en},
	urldate = {2019-01-25},
	publisher = {Elsevier},
	author = {Hedges, Larry V. and Olkin, Ingram},
	year = {1985},
	doi = {10.1016/C2009-0-03396-0},
}

@article{trafimow_null_2017,
	title = {Null hypothesis significance testing and {Type} {I} error: {The} domain problem},
	volume = {45},
	issn = {0732-118X},
	shorttitle = {Null hypothesis significance testing and {Type} {I} error},
	url = {http://www.sciencedirect.com/science/article/pii/S0732118X16301076},
	doi = {10.1016/j.newideapsych.2017.01.002},
	urldate = {2019-01-24},
	journal = {New Ideas in Psychology},
	author = {Trafimow, David and Earp, Brian D.},
	month = apr,
	year = {2017},
	pages = {19--27},
}

@article{morris_combining_2002,
	title = {Combining effect size estimates in meta-analysis with repeated measures and independent-groups designs.},
	volume = {7},
	issn = {1082-989X},
	url = {http://europepmc.org/abstract/med/11928886},
	doi = {10.1037/1082-989X.7.1.105},
	language = {eng},
	number = {1},
	urldate = {2019-01-24},
	journal = {Psychological methods},
	author = {Morris, S. B. and DeShon, R. P.},
	month = mar,
	year = {2002},
	pmid = {11928886},
	pages = {105--125},
}

@article{chinn_simple_2000,
	title = {A simple method for converting an odds ratio to effect size for use in meta-analysis},
	volume = {19},
	copyright = {Copyright © 2000 John Wiley \& Sons, Ltd.},
	issn = {1097-0258},
	url = {https://onlinelibrary.wiley.com/doi/abs/10.1002/1097-0258%2820001130%2919%3A22%3C3127%3A%3AAID-SIM784%3E3.0.CO%3B2-M},
	doi = {10.1002/1097-0258(20001130)19:22<3127::AID-SIM784>3.0.CO;2-M},
	language = {en},
	number = {22},
	urldate = {2019-01-24},
	journal = {Statistics in Medicine},
	author = {Chinn, Susan},
	year = {2000},
	pages = {3127--3131},
}

@book{cohen_statistical_1988,
	address = {Hillsdale, NJ},
	title = {Statistical power analysis for the behavioral sciences},
	publisher = {Erlbaum Associates},
	author = {Cohen, Jacob},
	year = {1988},
}

@article{wasserstein_asas_2016,
	title = {The {ASA}’s statement on p-values: context, process, and purpose},
	volume = {70},
	shorttitle = {The {ASA}’s statement on p-values},
	number = {2},
	journal = {The American Statistician},
	author = {Wasserstein, Ronald L. and Lazar, Nicole A.},
	year = {2016},
	pages = {129--133},
}

@article{wasserstein_moving_2019,
	title = {Moving to a {World} {Beyond} “p {\textbackslash}textless 0.05”},
	volume = {73},
	issn = {0003-1305},
	url = {https://doi.org/10.1080/00031305.2019.1583913},
	doi = {10.1080/00031305.2019.1583913},
	number = {sup1},
	urldate = {2019-03-21},
	journal = {The American Statistician},
	author = {Wasserstein, Ronald L. and Schirm, Allen L. and Lazar, Nicole A.},
	month = mar,
	year = {2019},
	pages = {1--19},
}

@article{corbett_developmental_2021,
	title = {Developmental effects in physiological stress in early adolescents with and without autism spectrum disorder},
	volume = {125},
	issn = {0306-4530},
	url = {https://www.sciencedirect.com/science/article/pii/S0306453020305382},
	doi = {10.1016/j.psyneuen.2020.105115},
	language = {en},
	urldate = {2021-02-26},
	journal = {Psychoneuroendocrinology},
	author = {Corbett, Blythe A. and Muscatello, Rachael A. and Kim, Ahra and Patel, Kunj and Vandekar, Simon},
	month = mar,
	year = {2021},
	pages = {105115},
}

@article{vandekar_robust_2019,
	title = {Robust spatial extent inference with a semiparametric bootstrap joint inference procedure},
	volume = {75},
	issn = {1541-0420},
	url = {https://onlinelibrary.wiley.com/doi/abs/10.1111/biom.13114},
	doi = {10.1111/biom.13114},
	language = {en},
	number = {4},
	urldate = {2021-11-30},
	journal = {Biometrics},
	author = {Vandekar, Simon N. and Satterthwaite, Theodore D. and Xia, Cedric H. and Adebimpe, Azeez and Ruparel, Kosha and Gur, Ruben C. and Gur, Raquel E. and Shinohara, Russell T.},
	year = {2019},
	pages = {1145--1155},
}

@article{kelley_effect_2012,
	title = {On effect size.},
	volume = {17},
	number = {2},
	journal = {Psychological methods},
	author = {Kelley, Ken and Preacher, Kristopher J.},
	year = {2012},
	pages = {137},
}

@book{achenbach_manual_2001,
	title = {Manual for the {ASEBA} school-age forms \& profiles : an integrated system of multi-informant assessment},
	isbn = {978-0-938565-73-4},
	shorttitle = {Manual for the {ASEBA} school-age forms \& profiles},
	url = {http://archive.org/details/manualforasebasc0000ache},
	language = {eng},
	urldate = {2025-08-25},
	publisher = {Burlington, VT : ASEBA},
	author = {Achenbach, Thomas M.},
	year = {2001},
}

@misc{zhang_semiparametric_2025,
	title = {Semiparametric {Confidence} {Sets} for {Arbitrary} {Effect} {Sizes} in {Longitudinal} {Neuroimaging}},
	copyright = {© 2025, Posted by Cold Spring Harbor Laboratory. The copyright holder for this pre-print is the author. All rights reserved. The material may not be redistributed, re-used or adapted without the author's permission.},
	url = {https://www.biorxiv.org/content/10.1101/2025.02.10.637497v1},
	language = {en},
	urldate = {2025-10-18},
	publisher = {bioRxiv},
	author = {Zhang, Xinyu and Liao, Kenneth and Seidlitz, Jakob and McHugo, Maureen and Avery, Suzanne N. and Huang, Anna and Alexander-Bloch, Aaron and Woodward, Neil and Heckers, Stephan and Vandekar, Simon},
	month = feb,
	year = {2025},
	doi = {10.1101/2025.02.10.637497},
}

@article{lombardo_big_2019,
	title = {Big data approaches to decomposing heterogeneity across the autism spectrum},
	volume = {24},
	copyright = {2019 The Author(s)},
	issn = {1476-5578},
	url = {https://www.nature.com/articles/s41380-018-0321-0},
	doi = {10.1038/s41380-018-0321-0},
	language = {en},
	number = {10},
	urldate = {2025-11-26},
	journal = {Molecular Psychiatry},
	author = {Lombardo, Michael V. and Lai, Meng-Chuan and Baron-Cohen, Simon},
	month = oct,
	year = {2019},
	note = {Publisher: Nature Publishing Group},
	pages = {1435--1450},
}

@article{kovacs_cdi_1985,
  title   = {The {Children}'s Depression Inventory ({CDI})},
  author  = {Kovacs, Maria},
  journal = {Psychopharmacology Bulletin},
  year    = {1985},
  volume  = {21},
  number  = {4},
  pages   = {995--998}
}

@article{feliciano_spark_2018,
  title={SPARK: A US cohort of 50,000 families to accelerate autism research},
  author={Feliciano, Pamela and Daniels, Amy M and Snyder, LeeAnne Green and Beaumont, Amy and Camba, Alexies and Esler, Amy and Gulsrud, Amanda G and Mason, Andrew and Gutierrez, Anibal and Nicholson, Amy and others},
  journal={Neuron},
  volume={97},
  number={3},
  pages={488--493},
  year={2018},
  publisher={Elsevier}
}





\label{lastpage}

\end{document}